\documentclass[aps, prfluids, noshowpacs,superscriptaddress,longbibliography]{revtex4}

\usepackage[english]{babel}
\usepackage{amsfonts}
\usepackage{verbatim}
\usepackage{amsmath}
\usepackage{amsthm}
\usepackage{amssymb}
\usepackage{graphicx}
\usepackage{subfig}
\usepackage{cancel}
\usepackage{physics}
\usepackage{siunitx}
\usepackage{xfrac}
\usepackage[usenames,dvipsnames]{color}
\usepackage{color,linegoal}
\usepackage[procnames]{listings}

\usepackage{setspace}

\newcommand{\Gammae} {\Gamma_\mathrm{e}}
\definecolor{myc}{RGB}{0,0,0}

\usepackage{natbib}

\begin{document}
\title{Ultra-low effective interfacial tension between miscible molecular fluids}

\author{Alessandro~Carbonaro}
\email{alessandro.carbonaro@umontpellier.fr}
\affiliation{Laboratoire Charles Coulomb (L2C), UMR 5221 CNRS-Universit\'{e} de Montpellier,
Montpellier, France}
\author{Luca~Cipelletti}
\affiliation{Laboratoire Charles Coulomb (L2C), UMR 5221 CNRS-Universit\'{e} de Montpellier,
Montpellier, France}
\author{Domenico~Truzzolillo}
\affiliation{Laboratoire Charles Coulomb (L2C), UMR 5221 CNRS-Universit\'{e} de Montpellier,
Montpellier, France}

\begin{abstract}
We exploit the deformation of drops spinning in a denser background fluid to investigate the effective interfacial tension (EIT) between miscible molecular fluids.
We find that for sufficiently low interfacial tension, spinning drops develop dumbbell shapes, with two large heads connected by a thinner central body. We show that this shape depends not only on the density and viscosity contrast between the drop and background fluids, but also on the fluid molecular structure, and hence on the stresses developing at their interface due to different molecular interaction. We systematically investigate the dynamics of dumbbell-shaped drops of water-glycerol mixtures spinning in a pure glycerol reservoir. By developing a model for the deformation based on the balance of the shear stress opposing the deformation, the imposed normal stress on the drop and an effective interfacial tension, we exploit the time evolution of the drop shape to measure the EIT. Our results show that the EIT in water-glycerol systems is orders of magnitude lower than that reported in previous experimental measurements, and in excellent agreement with values calculated via the phase field model proposed in [\emph{Phys. Rev. X 6, 041057, 2016}].
\end{abstract}

\maketitle

\section{Introduction}\label{sec_intro}

Interfacial tension between immiscible fluids is a well-defined, well-known quantity of paramount importance in a wide range of phenomena, from soft matter and material science to biophysics, oil recovery and multiphase flow \cite{rowlinson_molecular_2002}. By contrast, the presence of capillary stresses at the interface between miscible fluids is still debated and actively investigated. For miscible fluids, equilibrium thermodynamics states that interfacial tension should not exist, the equilibrium state being a homogeneous mixture of the fluids. However, transient capillary stresses between miscible fluids were first postulated in 1901 by D. Korteweg, who asserted that stresses due to density (or composition) gradients in a multifluid system could act as an effective interfacial tension (EIT) \cite{korteweg_kortewegpdf_1901}. Following his work, one can write the EIT, hereinafter denoted by $\Gammae$, similarly to the tension at equilibrium between immiscible fluids, i.e. by expanding the mixing free energy in even powers of the concentration gradient $\grad{\tilde{\varphi}}$ \cite{cahn_free_1958}. By considering only the first term of this expansion, $\Gamma_e$ can be written as:
\begin{equation}\label{Korteweg_Gamma}
\Gammae = \int_{-\infty}^{+\infty} \kappa(\tilde{\varphi}) (\grad{\tilde{\varphi}})^2 dz\,.
\end{equation}
Here $\tilde{\varphi}$ is the space-dependent volume fraction of one of the two fluids, $z$ is the coordinate orthogonal to the interface, and $\kappa(\tilde{\varphi})$ is the so-called Korteweg parameter, embedding the effect of the specific interaction between the fluids \cite{truzzolillo_off-equilibrium_2017,truzzolillo_nonequilibrium_2016}.
Clearly, $\Gammae$ tends to zero with time $t$, as diffusion smears out the interface, whose thickness increases, reducing $\grad{\tilde{\varphi}}$ with $t$ \cite{cicuta_capillary--bulk_2001}. Such a transient, out-of-equilibrium interfacial tension has been invoked in literature to rationalize the behaviour of miscible fluids at short times, before they are fully mixed, and several works tried to elucidate the role of stresses at miscible boundaries, both theoretically \cite{vorobev_shapes_2020, antanovskii_microscale_1996,bier_nonequilibrium_2015,vorobev_phase-field_2016,chen_miscible_1996,chen_miscible_2002} and experimentally.
Among the strategies adopted to measure the EIT between miscible fluids, the most recent ones leverage on the study of hydrodynamic instabilities \cite{truzzolillo_nonequilibrium_2016, shevtsova_two-scale_2016}, on light scattering experiments probing capillary waves \cite{cicuta_capillary--bulk_2001, may_capillary-wave_1991} and on the observation of the shape of drops and threads under an external forcing \cite{pojman_evidence_2006,carbonaro_spinning_2019,gowda_v_effective_2019}. Despite of this effort, the magnitude and even the very existence of EIT between simple molecular fluids is still debated and mostly unclear.

One technique to measure very low interfacial tensions ($10^{-3}$ - $10^{-2}\,\si{\milli\newton/\meter}$ \cite{bamberger_effects_1984,liu_concentration_2012}) is spinning drop tensiometry (SDT), which is based on the observation of drop shapes. In an SDT experiment a drop is injected in a denser background fluid contained in a cylindrical capillary. When the capillary is spun, the drop elongates on the axis of rotation due to centrifugal forces. Following the drop shape by means of video imaging, one can then measure the interfacial tension between the drop and the background fluid. In the case of immiscible fluids, for which SDT was initially conceived by Vonnegut \cite{vonnegut_rotating_1942}, one typically measures the equilibrium shape of the drop, which is dictated by the balance between surface tension and centripetal forces. The interfacial tension $\Gamma$ is then obtained through the Vonnegut equation \cite{vonnegut_rotating_1942}:
\begin{equation}\label{vonnegut}
	\Gamma = \frac{\Delta\rho \omega^2 r^3}{4}\,,
\end{equation}
where $\Delta\rho$ is the density difference between the background and drop fluids, $\omega$ the angular velocity and $r$ the equilibrium radius of the drop. A second possibility is to characterize the time evolution of the drop after a sudden rotational speed jump. Recently, we have employed this technique to study the elongation dynamics of drops, both in miscible and immiscible background fluids \cite{carbonaro_spinning_2019}, showing that the drop dynamics towards an equilibrium state are characterized by a relaxation time fully determined by i) the viscosity of the fluids, ii) the drop size and iii) the interfacial tension.
In the past few decades, SDT experiments aiming at measuring equilibrium states have been performed by several groups to investigate the presence of an EIT between miscible fluids, either close \cite{pojman_evidence_2006} or far from a spinodal decomposition of the fluids \cite{zoltowski_evidence_2007, petitjeans_petitjean_1996}. Unfortunately, in the case of fully miscible molecular liquids, such as water and glycerol, diffusion hampers the measurement of stationary states and literature data are conflicting, sometimes even in experiments by the same authors \cite{legendre_instabilites_2003, petitjeans_petitjean_1996}. Indeed, when $\Gamma$ is negligible a stationary state is never attained, as we showed for one specific pair of miscible fluids with small compositional mismatch, namely a drop of a water-glycerol mixture (5 $\%$ wt H$_2$O) spinning in pure glycerol. As a result, the question of whether an EIT exists or not in such a case has not been settled yet.

SDT experiments at low $\Gamma$ are furthermore complicated by the fact that drops do not always maintain a simple ellipsoidal shape. Even for immiscible fluids, for sufficiently low interfacial tensions, they can develop a ``dumbbell", or ``dog-bone", shape consisting in two large heads connected by a thinner central body, as reported in the case of water-hydrocarbon-surfactant systems with $\Gamma<$ 10 $\si{\micro\newton/\meter}$ \cite{manning_interfacial_1977}. In this case a satisfactory explanation of the phenomenon is still lacking. More recently \cite{zoltowski_evidence_2007}, such shapes have been also observed in miscible fluids and have been attributed to the effect of perturbation due to viscous secondary flows in finite reservoirs of rotating fluids. Such an effect, till now unexplored, is the focus of the present work.

\begin{figure}
	\includegraphics[width = \linewidth]{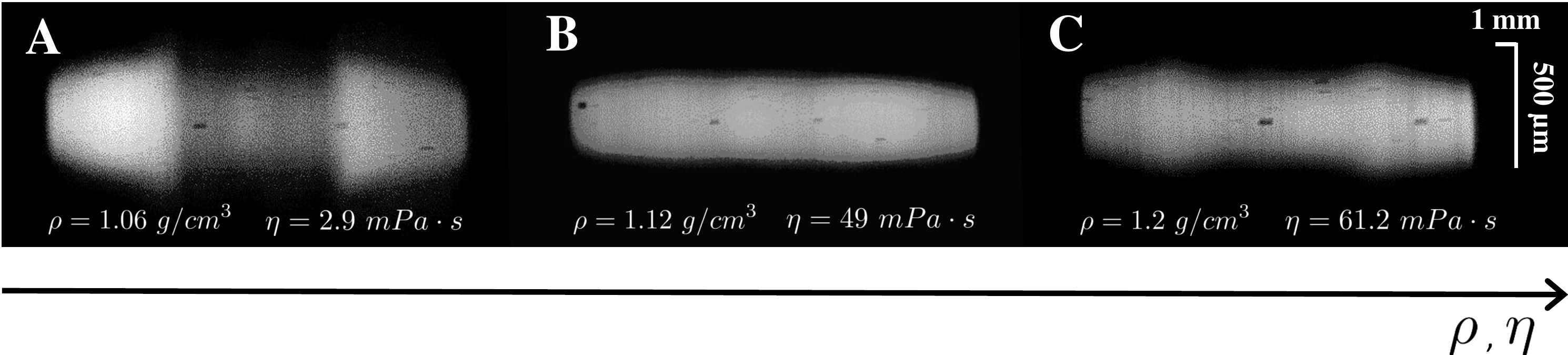}
	\caption{Dyed drops containing 75\% H\textsubscript{2}O-25\% glycerol (Panel A),  TEG (Panel B) and 20\% H\textsubscript{2}O-80\% glycerol (Panel C). All drops spin in a reservoir of pure glycerol. Images are optically compressed in the horizontal direction and expanded in the vertical direction to improve resolution, as detailed in the main text (section \ref{sec_materials_methods}).}
	\label{Carabinieri}
\end{figure}

To set the scene, Fig. \ref{Carabinieri} shows examples of such dumbbell shapes for drops rotating in a pure glycerol reservoir. Drops in panels A and C are composed of mixtures of water and glycerol, respectively with water mass concentration $c_w = 0.75$ and $c_w = 0.20$, whereas the drop in panel B is made of triethylene glycol (TEG). All drops are fully miscible with the glycerol background.
Strikingly, Fig. \ref{Carabinieri} shows that the development of a dumbbell shape depends not only on the density and viscosity contrast with respect to the background fluid, but also on the molecular structure of the fluids. Indeed, in absence of any interfacial stresses, the TEG drop should have an intermediate shape between drops in panels A and C, since it has intermediate density and viscosity.
This is evidently not the case: consequently, Fig. \ref{Carabinieri} cannot be explained only by means of hydrodynamic arguments. Thus, characterizing the evolution of such drop shapes appears as a promising strategy to measure the effect of interfacial stresses between the drop and background fluids. In this work, we tackle this challenging task and investigate experimentally the time evolution and the origin of dumbbell-shaped drops by systematically varying the composition of the drop fluid in a series of SDT experiments.
Furthermore, we exploit fluorescent drops to track the time evolution of the full concentration profile of the fluids in the capillary, instead of simply measuring the intensity profile of the collected light. We model the temporal dynamics of the drop shape by balancing the normal stress imposed on the drop surface, the shear stress opposing the deformation and the effect of an EIT, and we exploit the deformation dynamics to measure the effective interfacial tension between miscible molecular fluids.

The rest of the work is organized as follows. In section \ref{sec_materials_methods} we present the setup and materials employed, and elucidate the procedure to extract the concentration profile of the fluids in the capillary. In section \ref{sec_results_discussion} we present the data on the deformation dynamics of drops, discussing our results in the light of a model allowing one to measure $\Gamma_e$.
Finally, in section \ref{sec_conclusions} we make some concluding remarks and summarize the key results of our work.

\section{Materials and methods} \label{sec_materials_methods}
Glycerol ($\geq$99.5\% wt) was purchased from Sigma Aldrich and used as received. Water-glycerol mixtures were prepared using Milli-Q ultra pure water, with densities $\rho$ and viscosities $\eta$ reported in Table \ref{tab:table1} as a function of water mass fraction $c_w$, for T $= 25.0 \pm 0.5 ^\circ$C. The water mass fraction was determined via rheological measurements as detailed in Table \ref{tab:table1}. Fluorescein (disodium salt) was purchased from Merck KGaA and dissolved in the drop fluids at two concentrations, $\num{2e-3}$ wt/wt and $10^{-3}$ wt/wt, for two independent sets of measurements as detailed later.
Experiments were performed with a Kr\"{u}ss spinning drop tensiometer at $25.0 \pm 0.5 ^\circ$C, with rates of rotation ranging from $6000$ rpm to $15000$ rpm, so that buoyancy could be neglected. All drops were injected with a $1 \si{\micro\liter}$ syringe in a capillary with an internal diameter of $3.25$ $\si{\milli\meter}$, prefilled with glycerol before each experiment. The time between the injection and the beginning of the rotation was typically $10$ - $15$ seconds. After each measurement, the fluids were replaced and a fresh new drop was injected in the capillary.
To image the drops, the tensiometer is equipped with a blue LED with dominant wavelength of 469 $\si{\nano\meter}$ for illumination and a CMOS camera (Toshiba Teli BU406M) for imaging. Since the drops become very elongated, we use as an objective two cylindrical lenses (Newport CKX17-C) that expand the field of view in the horizontal direction $x$ and compress it in the vertical direction $y$. As described in \cite{carbonaro_spinning_2019}, the resulting magnification in the $x$ (horizontal) direction is $M_x=0.3$, while the magnification in the $y$ (vertical) direction is $M_y = 3.36$. The different horizontal and vertical magnification allows one to follow the dynamics of very elongated drops along almost all the capillary length (5 cm), while gaining at the same time in accuracy along the vertical direction. Following the magnification stage, a blue light filter in front of the CMOS eliminates the blue background light.
\begin{table}
	\begin{center}
		\centering
		\caption{\label{tab:table1}}
		\begin{ruledtabular}
			\begin{tabular}{ccc}
				Liquid & $\rho$ (g/cm$^3$) \footnote{Densities of water-glycerol mixtures were obtained from tabulated values \cite{noauthor_physical_1963} correspondent to mixtures having the measured zero-shear viscosities.}& $\eta$ (mPa s)\footnote{The viscosities of water-glycerol mixtures and pure glycerol were measured performing steady rate rheology experiments using a stress-controlled AR 2000 rheometer (TA Instruments) with a steel cone-and-plate geometry (cone diameter = 50 mm, cone angle = 0.0198 rad). No dependence of the viscosity on the shear rate was observed as all samples showed pure Newtonian response.}\\
				\hline
				Glycerol\footnote{The viscosity of the glycerol used as the background fluid is lower than that tabulated for anhydrous glycerol \cite{noauthor_physical_1963} because of the unavoidable adsorption of water from the atmosphere.} ($c_w\leq$0.02) & 1.26 $\pm$ 0.01 & 800.0 $\pm$ 0.1\\
				$c_w$=0.25& 1.19 $\pm$ 0.01 & 33.8 $\pm$ 0.1\\
				$c_w$=0.45& 1.12 $\pm$ 0.01 & 9.0 $\pm$ 0.1\\
				$c_w$=0.70& 1.07 $\pm$ 0.01 & 2.8 $\pm$ 0.1\\
				$c_w$=0.75& 1.06 $\pm$ 0.01 & 2.3 $\pm$ 0.1\\
				$c_w$=0.90& 1.02 $\pm$ 0.01 & 2.2 $\pm$ 0.2\\
				Water\footnote{The viscosity and the density of water have been
					measured using a rolling-ball Anton Paar Lovis 2000ME
					microviscosimeter and DMA 4500M densimeter, respectively.} & 0.996 $\pm$ 0.001& 0.89 $\pm$ 0.01\\
			\end{tabular}
		\end{ruledtabular}
	\end{center}
\end{table}

\subsection{Concentration profiles}\label{section_inversion}
\begin{figure}
	\centering
	\subfloat[\label{evolution_Intensity}]{\includegraphics[height = 5.75cm]{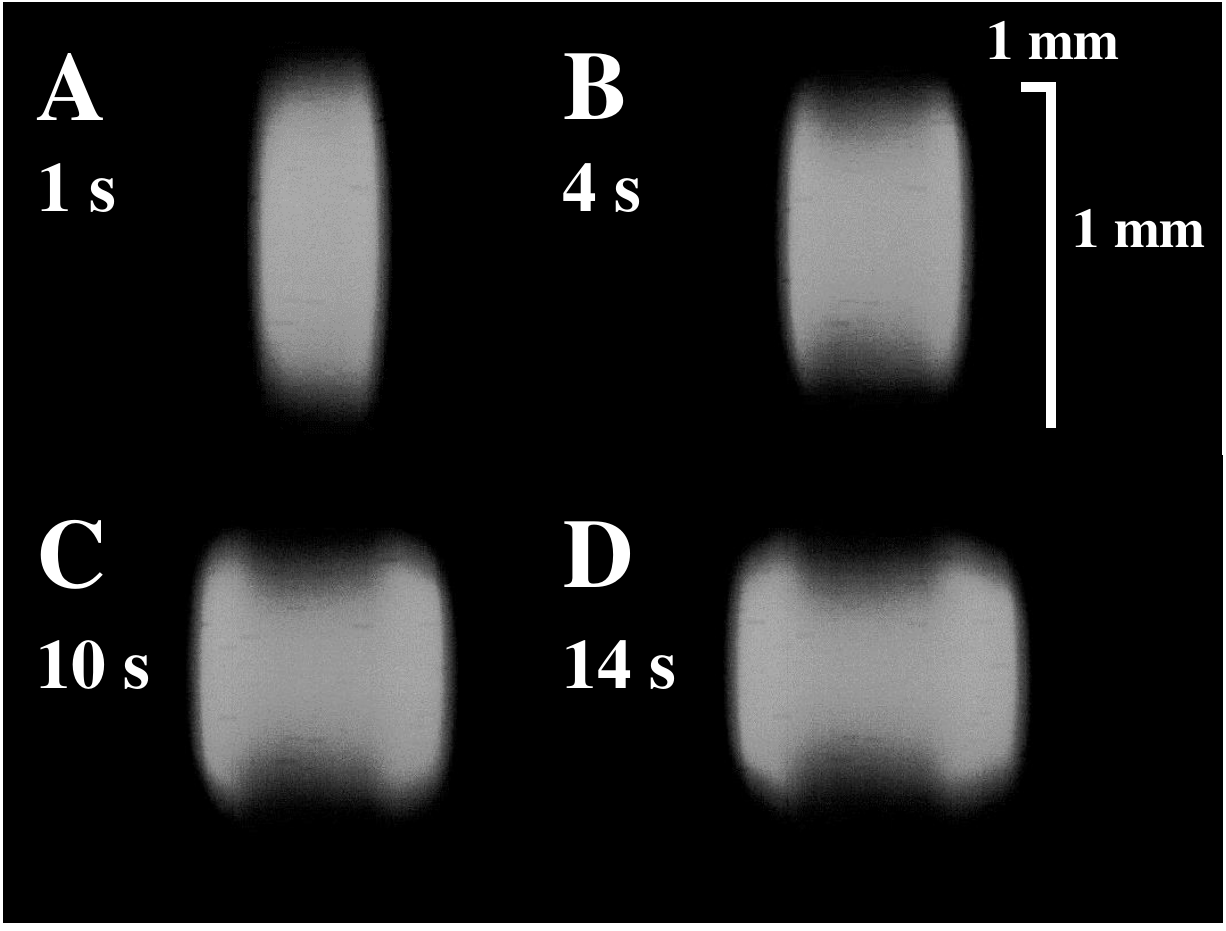}}
	\hfill
	\subfloat[\label{evolution_phi}]{\includegraphics[height = 5.75cm]{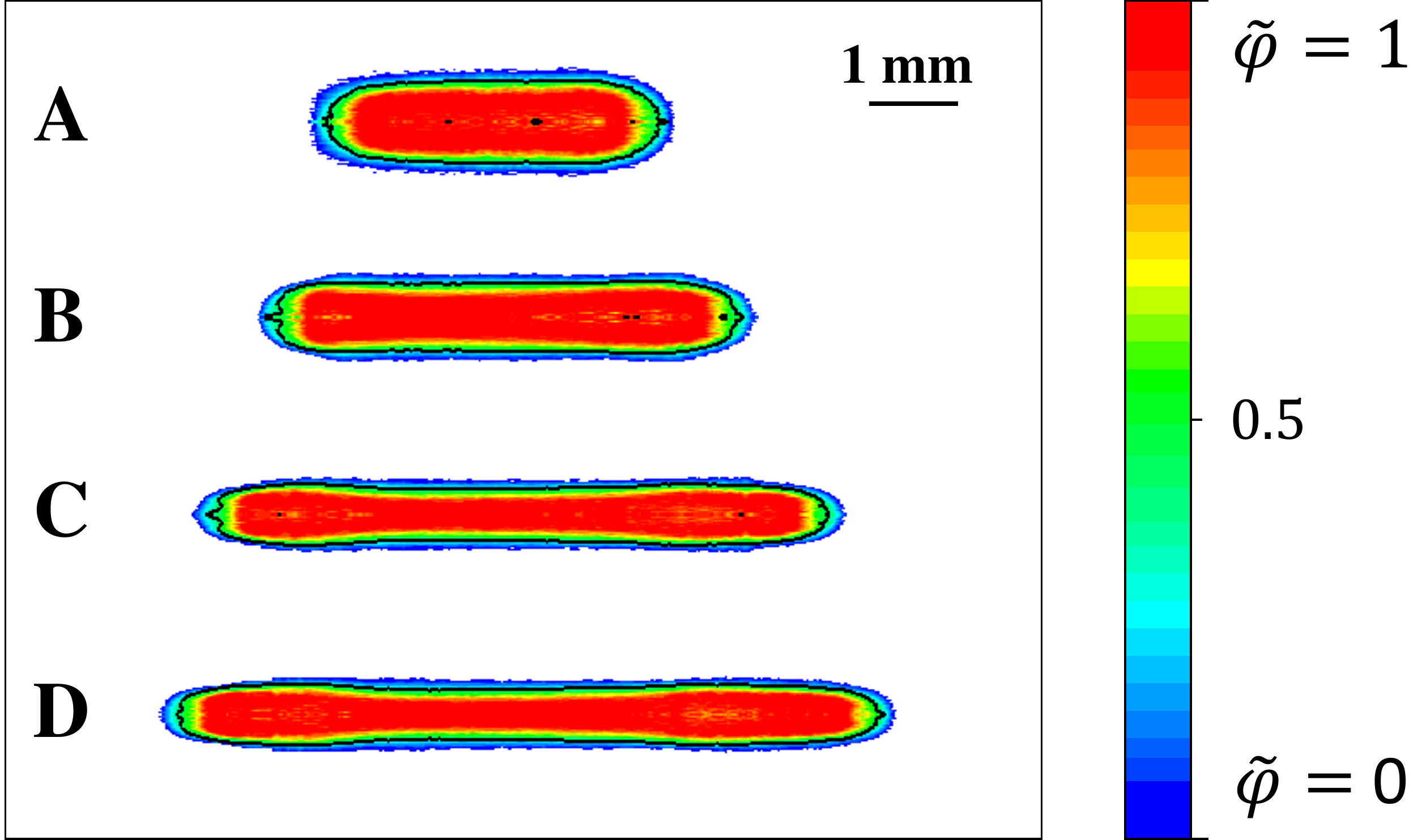}}
	\caption{(a) Time evolution of the fluorescence intensity recorded for a typical drop of pure water in a pure glycerol background. Note the different magnification in the horizontal and vertical directions. (b) Concentration profiles reconstructed from the fluorescence intensity images of panel (a), as detailed in the text. In panel (b), the scale is the same in the horizontal and vertical directions. The black line represents the drop surface, defined as the set of points where $\tilde{\varphi}(x,r)$=0.5.}	
\end{figure}

An example of the optically compressed images of the drops obtained with the experimental setup is shown in Fig. \ref{evolution_Intensity}, displaying the typical time evolution of a drop of pure water spinning in pure glycerol.
A further step is required to follow more precisely the evolution of the interface. To this end, we extract the concentration profile of the drop fluid from the fluorescence intensity, in order to define precisely the drop boundary. Note that, strictly speaking, the fluorescence signal comes from the spatial distribution of fluorescein, not that of the drop fluid. However, over the time scale of our experiments ($\sim$ 10 $\si{\second}$) we assume the concentration profile of fluorescein to closely follow the one of the drop fluid, since there is not enough time to develop an appreciable difference in distribution of the two components. To support this claim, we compare the diffusion coefficients of water (the drop fluid) and of fluorescein in the background glycerol. The self diffusion coefficient of water is $D_{w} = \num{1.025e-9}\,\si{\meter^2/\second}$ and the diffusion coefficient of water in glycerol is $D_{wg} = \num{1.4e-11}\,\si{\meter^2/s}$ \cite{derrico_diffusion_2004-1}. On the other hand, the diffusion coefficient of fluorescein in water is $D_{fw}=\num{6.4e-10}\,\si{\meter^2/\second}$ \cite{harrison_micro-fluidic_1998}. We estimate the diffusion coefficient of fluorescein in glycerol as $D_{fg}\simeq D_{fw} \frac{D_{wg}}{D_{w}} \simeq \num{8.7e-12}\,\si{\meter^2/\second}$. Therefore, the difference between the distances over which water and fluorescein may diffuse over the time scale of our experiments is $l_w-l_f\approx 3 \,\si{\micro\meter}$, much smaller than the resolution with which we measure the drop shape, which we determine to be few tens of $\si{\micro\meter}$. Consequently, we can safely assume the concentration profile of the fluorophore to represent well that of the drop fluid. Furthermore, by changing the concentration of fluorophore over more than one decade (from $\num{1e-4}$ wt/wt to $\num{2.5e-3}$ wt/wt), we tested that the concentration of fluorescein is directly proportional to the light intensity collected.

These considerations, together with an additional symmetry argument, allow linking the intensity of the collected light to the $x$-dependent radial concentration profile of the drop fluid. Exploiting the cylindrical symmetry of the drops, one can move from a bidimensional intensity image to a three-dimensional concentration map.

\begin{figure}
	\includegraphics[width = 0.6\linewidth]{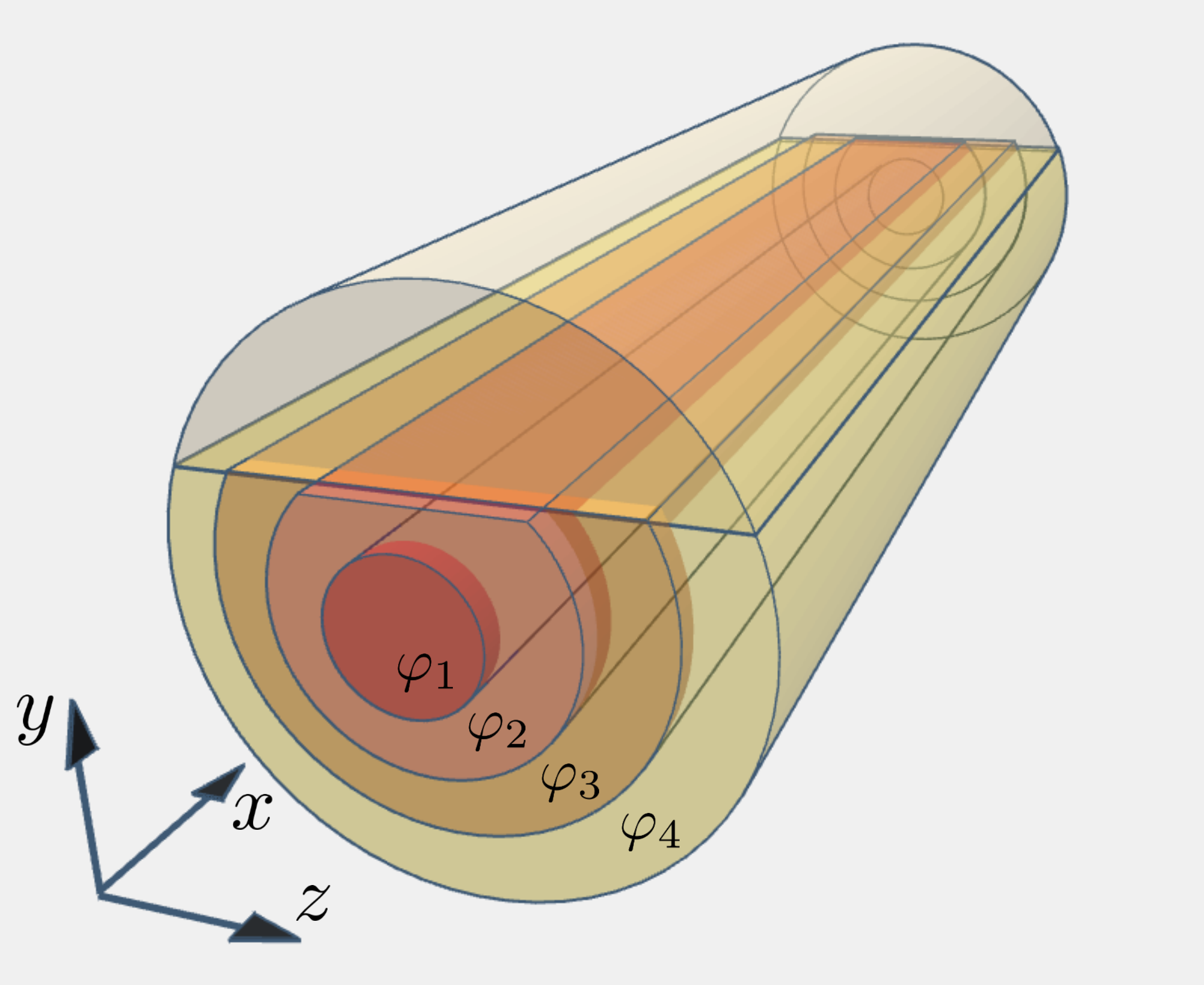}
	\caption{Scheme of the drop for the inversion routine to retrieve the concentration profile from the fluorescence intensity. We typically use $N = 204$ shells; for clarity, only the first four shells are shown here.}
	\label{shells}
\end{figure}

In order to do so, we divide the drop into $N$ concentric cylindrical shells of radius $r_i$ and constant thickness $dr$, each of them having a local value of water volume fraction $\tilde{\varphi}_i$ (see Fig. \ref{shells}). The fluorescence intensity at a given horizontal coordinate $x$ and vertical coordinate $y$ is be given by the sum over the contributions of each shell, weighted by a geometric factor $c_i$ proportional to the length of the chord that traverses such shell at coordinate $y$, along the $z$ direction, i.e. the line of sight.

Accordingly, the intensity distribution reads:
\begin{equation}\label{inversionEq}
	I(x_k,y_j) = \sum_{i=1}^{N} c_i(y_j) \tilde{\varphi}_i(x_k)\,,
\end{equation}
where the indexes $k$ and $j$ have been introduced to account for the discretization of $I$ on the pixel grid of the CMOS camera.
For a given $x_k$, one can recast the problem in the matrix form:
\begin{equation}\label{matrixForm}
	\mathbf{I}=\mathbf{C}\mathbf{\tilde{\varphi}}\,,
\end{equation}
with $\mathbf{I} = \{I_j\}$, $\mathbf{C} = \{c_{ji}\}$ and $\mathbf{\tilde{\varphi}} = \{\tilde{\varphi}_i\}$, and $I_j \equiv I(x_k,y_j)$, $c_{ji} \equiv c_i(y_j)$ and $\tilde{\varphi}_i\equiv \tilde{\varphi}_i(x_k)$.
For each $k$, the solution is thus $\mathbf{\tilde{\varphi}}=\mathbf{C}^{-1}\mathbf{I}$. The inversion of the matrix $\mathbf{C}$ is sped up by exploiting the symmetry of the drop around the longitudinal axis, which translates into the condition that $\mathbf{C}$ is lower-diagonal.
By solving Eq. \ref{matrixForm} for all $x_k$ of interest, one obtains the concentration map of the the drop fluid in the capillary, as shown in Fig. \ref{evolution_phi}. We use the maps to define the drop surface as the set of points where $\tilde{\varphi}(x,r)$=0.5, with $\tilde{\varphi}(x,r)$ normalized to unity in the region of the drop close to the axis of rotation. The amplitude $h$ of the drop deformation is defined as the difference between the maximum radius of the drop, close to the tips, and the minimum radius, at the center of the drop (see Fig. \ref{model} below).

\section{Results and discussion}\label{sec_results_discussion}
\begin{figure}
	\centering
	\subfloat[\label{A_t}]{\includegraphics[width = 0.5\linewidth]{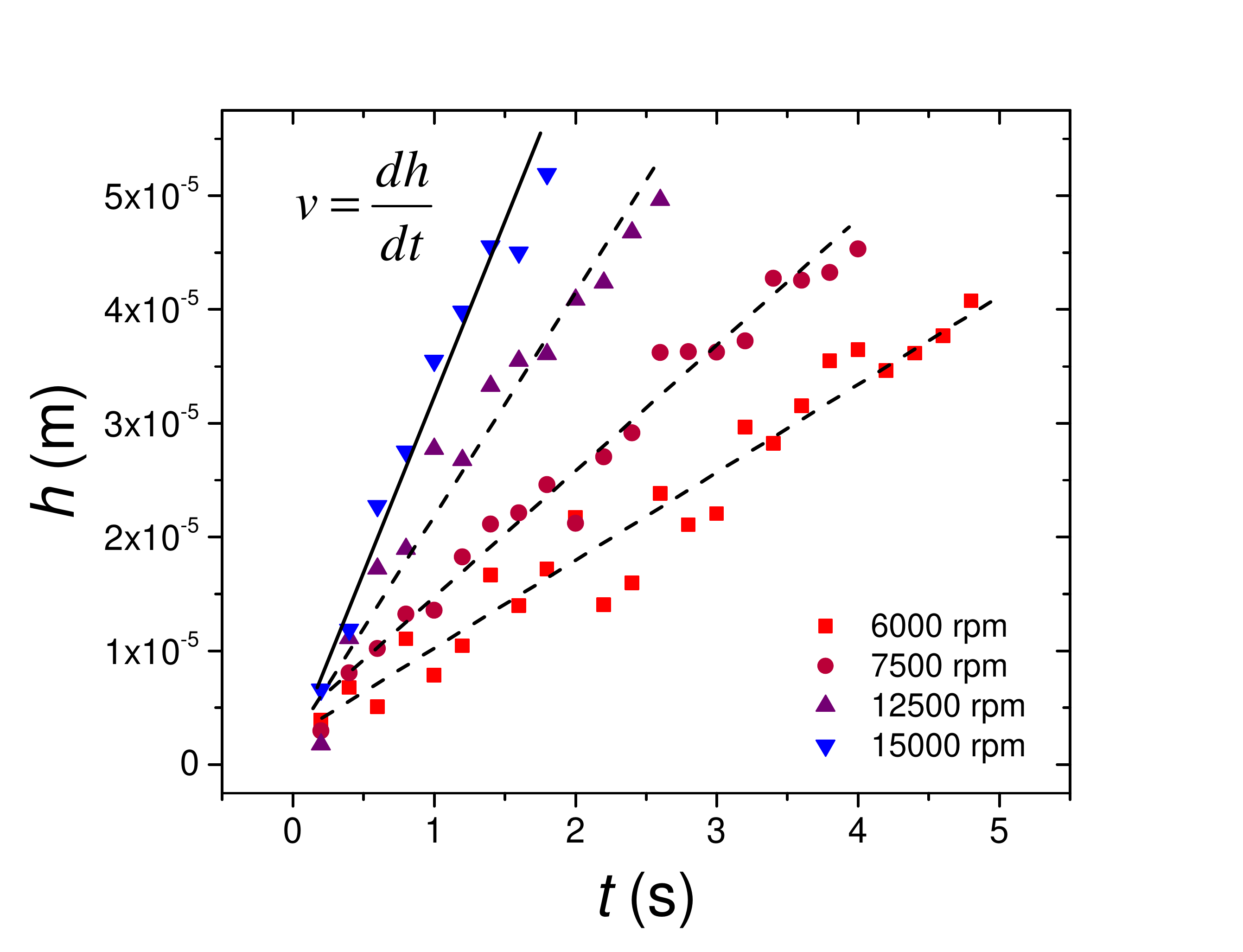}}
	\hfill
	\subfloat[\label{v_forcing}]{\includegraphics[width = 0.5\linewidth]{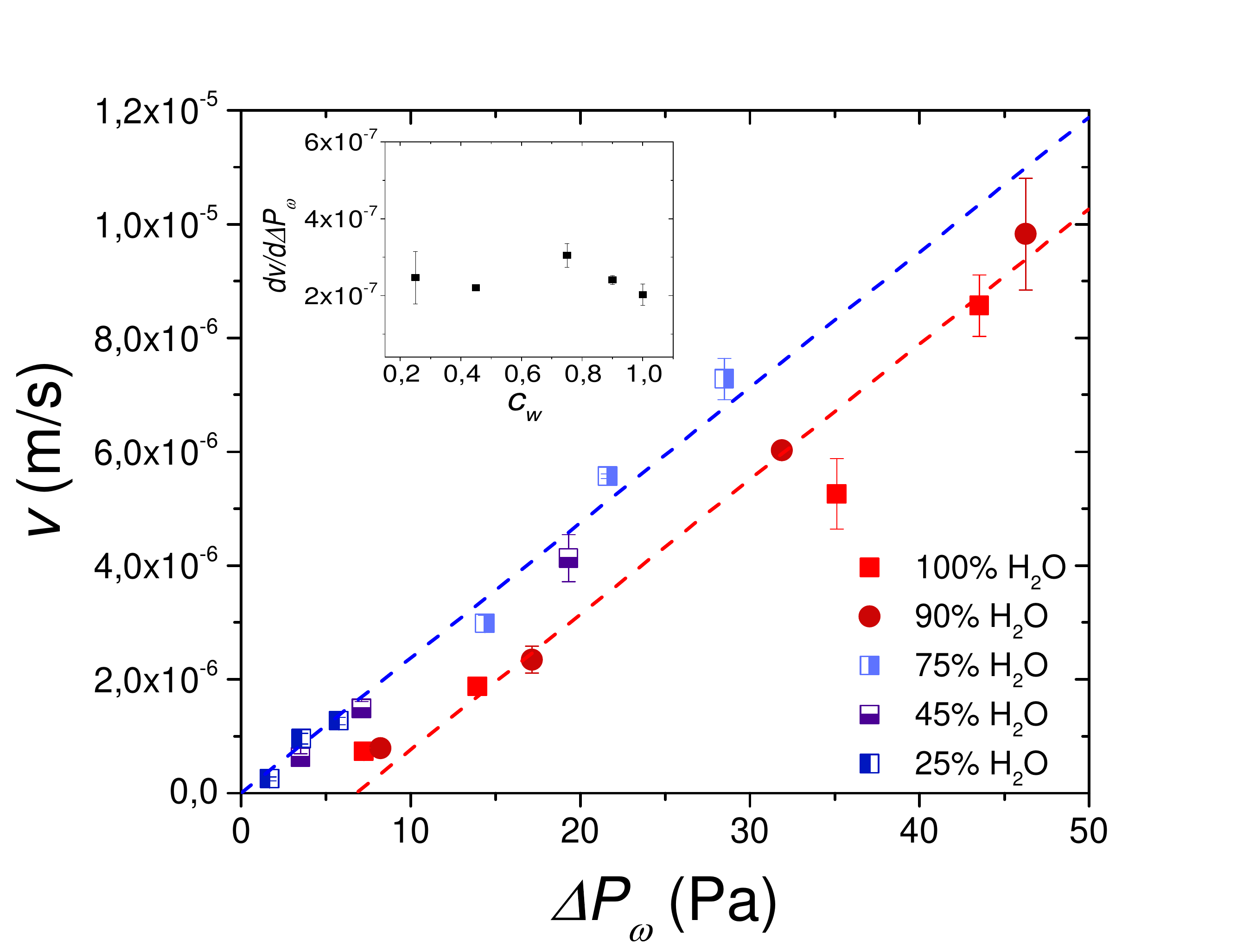}}
	\caption{(a) Time evolution of the deformation amplitude $h$ for a drop of pure water in a pure glycerol background, at different $\omega$ as shown by the labels. (b) Deformation velocity as a function of the centripetal forcing, for various water mass fractions in the drop. The inset shows the slope of the velocity versus $\Delta P_\omega$, obtained by fitting independently datasets at a given $c_w$ with a straight line.}	
\end{figure}

We characterize the evolution of the drops towards a dumbbell shape by observing the time evolution of the deformation amplitude $h$, varying systematically the drop composition.
Figure \ref{A_t} shows the time evolution of $h$ for a drop of pure water in pure glycerol, for various rotational speeds. Time $t=0$ corresponds to the onset of the deformation, shortly after the start of rotation. At short time the deformation amplitude increases linearly with time, with a velocity $v=\frac{dh}{dt}$ that depends on the rotation speed. This dependence on $\omega$ stems from the fact that spinning drops are subject at the head of the drop to a rotation-induced pressure jump $\Delta P_\omega = \frac{1}{2} \Delta\rho \omega^2 r^2$ \cite{carbonaro_spinning_2019, lister_time-dependent_1996}, where $\Delta\rho$ is the density difference between the denser background fluid and the drop and $r$ the drop radius. This pressure jump $\Delta P_\omega$ induces the drop elongation in first place, but it is also responsible for a secondary flow of the background fluid that leads to the dumbbell shape, as we shall detail below. Figure \ref{v_forcing} reports the radial deformation velocity as a function of $\Delta P_\omega$, for various concentration of water in the drops.

It is worth underlining two features. First, the velocity of the radial deformation is approximately a linear function of $\Delta P_\omega$, with a slope that is essentially independent of the concentration of water in the drop, as shown in the inset of Fig. \ref{v_forcing}. Second, even before trying to rationalize the dependence of $v$ on $\Delta P_\omega$, it appears clearly that the data may be divided in two main families.
All data for drops with a mass concentration of water $c_w \le 0.75$ are compatible with a straight line through the origin. By contrast, data from drops with $c_w = 0.9$ and $c_w = 1$ show a different behaviour in that a linear fit of $v(\Delta P_\omega)$ displays a negative intercept with the $v$-axis. This is counterintuitive if we were to neglect interfacial tension: drops containing a larger fraction of water, being less viscous, should be deformed more rapidly for a given centripetal forcing. Thus, the negative intercept of data for $c_w \geq 0.9$ strongly suggests a non-vanishing EIT for these systems.
As a first step towards the modelling of our experiments, we perform a linear fit of two families of data in Fig. \ref{v_forcing}, for water concentrations up to $c_w = 0.75$ and above $c_w = 0.9$, respectively:
\begin{equation}\label{stat_model}
	v = A \Delta P_\omega + B\,,
\end{equation}
finding an $R^2$ value of 0.98 and 0.97 respectively, keeping the slope $A$ the same for the two data families. The physical meaning of the terms $A$ and $B$ will be detailed later; however, we anticipate that $B$ depends on the concentration $c_w$ of water in the drop and hence on the EIT. It is thus crucial to perform a statistical analysis of the data of Fig. \ref{v_forcing} in order to assess whether the difference in the offset $B$ between data below and above $c_w = 0.9$ (semifilled and solid symbols, respectively) is statistically significant, or just due to experimental noise.

We perform a t-test \cite{m_r_spiegel_statistics_2008, press_numerical_1996} on the difference between term $B$ of the fit for the two data families, for $c_w\leq 0.75$ and $c_w \geq 0.9$. As detailed in \cite{noauthor_supplemental_nodate}, for the data of Fig. \ref{v_forcing} the Student t-distribution yields a value of the standardized variable $t = 6.43$. This is much larger than $t_{0.995} = 2.95$, the edge of the $1\%$ confidence interval for a two-tailed t-distribution with $N_1+N_2-2 = 15$ degrees of freedom, where $N_1 = 9$ and $N_2 = 8$ are the number of data points in the two datasets. Hence we conclude that the difference between the $B$ parameter for the two datasets in Fig. \ref{v_forcing} is statistically significant, with a confidence level greater than 99\%.

\subsection{Model of the radial deformation and EIT}
\begin{figure}
	\includegraphics[width = \linewidth]{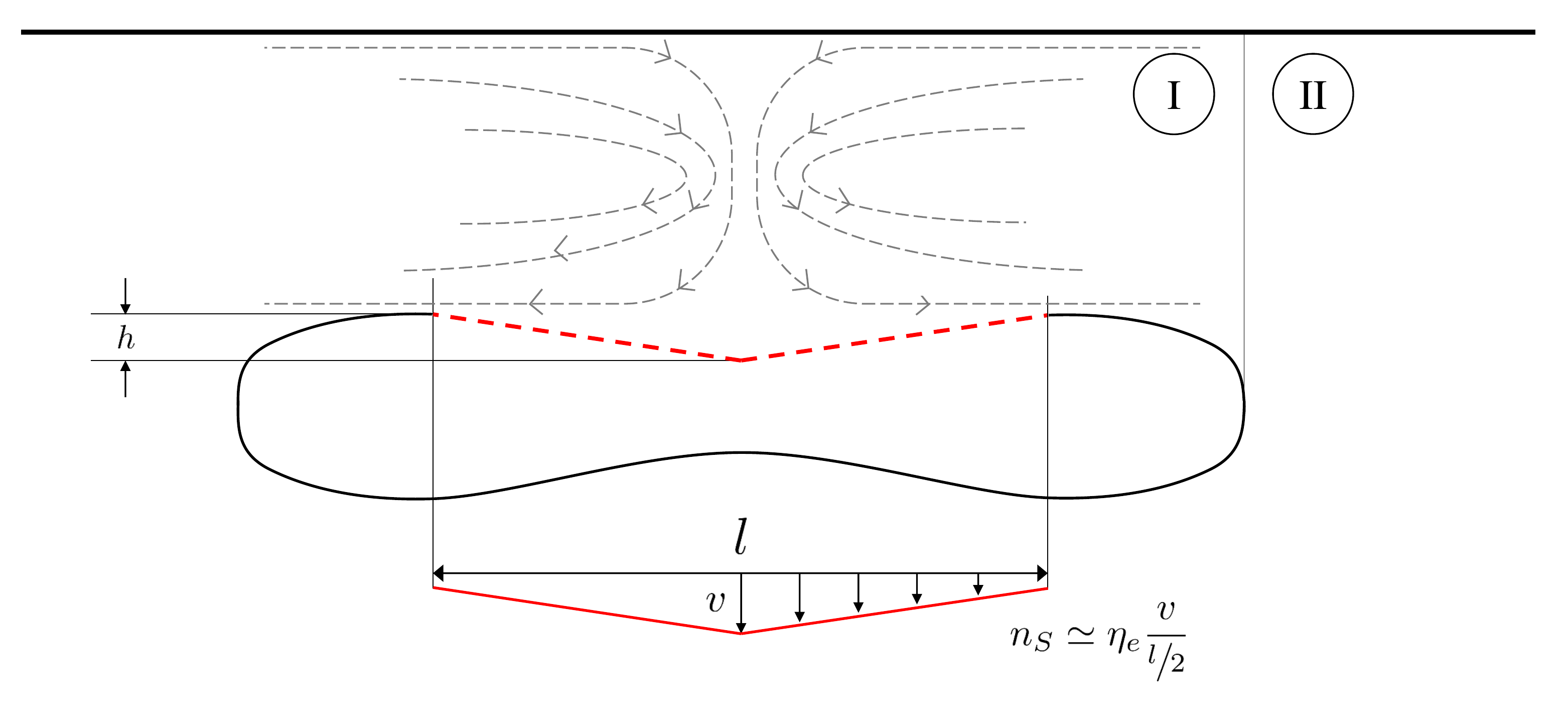}
	\caption{Scheme of the drop deformation. The dashed arrows show the secondary flow of the background fluid induced by the capillary rotation.}
	\label{model}
\end{figure}

Having checked that the effect of the EIT on the data of Fig. \ref{v_forcing} is statistically significant, we propose a simple model to rationalize the data and extract from them $\Gamma_e$. The model is based on the balance of all sources of stress on the drop interface at the onset of the deformation.
Similarly to the approach by Lister and Stone in \cite{lister_time-dependent_1996}, we write en equilibrium equation for the stresses on the drop surface, at the center of the drop:
\begin{equation}\label{balanceP}
	n_E = n_S + n_L\,,
\end{equation}
where $n_E$ is the normal stress inducing the deformation and $n_S$ and $n_L$ are the normal stresses opposing the deformation, with $n_S$ the shear stress arising from the motion of the drop and background fluids and $n_L$ a Laplace-like term that takes into account the effect of the EIT. 
Since in our experiments the viscosity $\eta_e$ of the external background fluid is much higher than the viscosity $\eta_d$ of the drop fluid, the shear stress can be approximated as $n_S\simeq 2\eta_e v/l$, where $l$ is the distance over which the radial deformation develops as shown in Fig. \ref{model}.
The Laplace-like term $n_L$ is estimated as the difference between the Laplace pressure jump at the heads of the drop and that at the central body. At the onset of the deformation, the drop shape is well described by a cylinder with radius $r_0$ capped by two hemispheres. Consequently, the hemispherical heads and the cylindrical body will be characterised by a pressure jump with respect to the background fluid of $\sfrac{2\Gammae}{r_0}$ and $\sfrac{\Gammae}{r_0}$ respectively, leading to $n_L = \sfrac{\Gammae}{r_0}$, where $\Gammae$ is the EIT.

The normal stress $n_E$ inducing the deformation and arising from the external forcing on the drop deserves a further discussion.
When investigating the limits of the SDT technique, as early as in 1982, Currie and Van Nieuwkoop observed that in any spinning capillary the background fluid is not at rest, but rather flows towards the axis of the capillary at the center of the drop, thus inducing an extra normal stress on the drop surface \cite{currie_buoyancy_1982, manning_interfacial_1977}. This secondary recirculating flow pushing on the drop surface originates from the jump in centripetal pressure at the drop head, between regions I and II in Fig. \ref{model}. The origin of this jump is easily understood by recalling that the hydrostatic pressure induced by the rotational acceleration is proportional to the fluid density, which is smaller in the drop as compared to the background fluid.
By numerically solving the Navier-Stokes equations, we verified that this secondary flow gives rise to a velocity field with a radial component directly proportional to $\Delta P_\omega$ \cite{noauthor_supplemental_nodate}. Since the external forcing on the drop originates from the secondary recirculating flow and the latter is proportional to $\Delta P_\omega$, we write $n_E=\alpha \Delta P_\omega$ with $\alpha$ a positive constant.


Equation \ref{balanceP} can then be rewritten as
\begin{equation}
	\alpha \Delta P_\omega -\frac{2 \eta_e v}{l} - \frac{\Gammae}{r_0} = 0\,,
\end{equation}
which yields for the radial deformation velocity
\begin{equation}\label{velocity}
v = \frac{\alpha l}{2 \eta_e}\Delta P_\omega - \frac{\Gammae l}{2 r_0 \eta_e}\,,
\end{equation}
i.e. the linear form introduced empirically  in Eq. \ref{stat_model}.
Note that for miscible fluids $\Gammae$ decreases over time, such that Eq. \ref{velocity} holds only for short times after the onset of the radial deformation, well before diffusion smears out the interface. For this reason, we measure the dynamics of the drop deformation only for a few seconds, before diffusion becomes significant.

\begin{figure}
	\centering
	\subfloat[\label{mastercurve}]{\includegraphics[width = 0.45\linewidth]{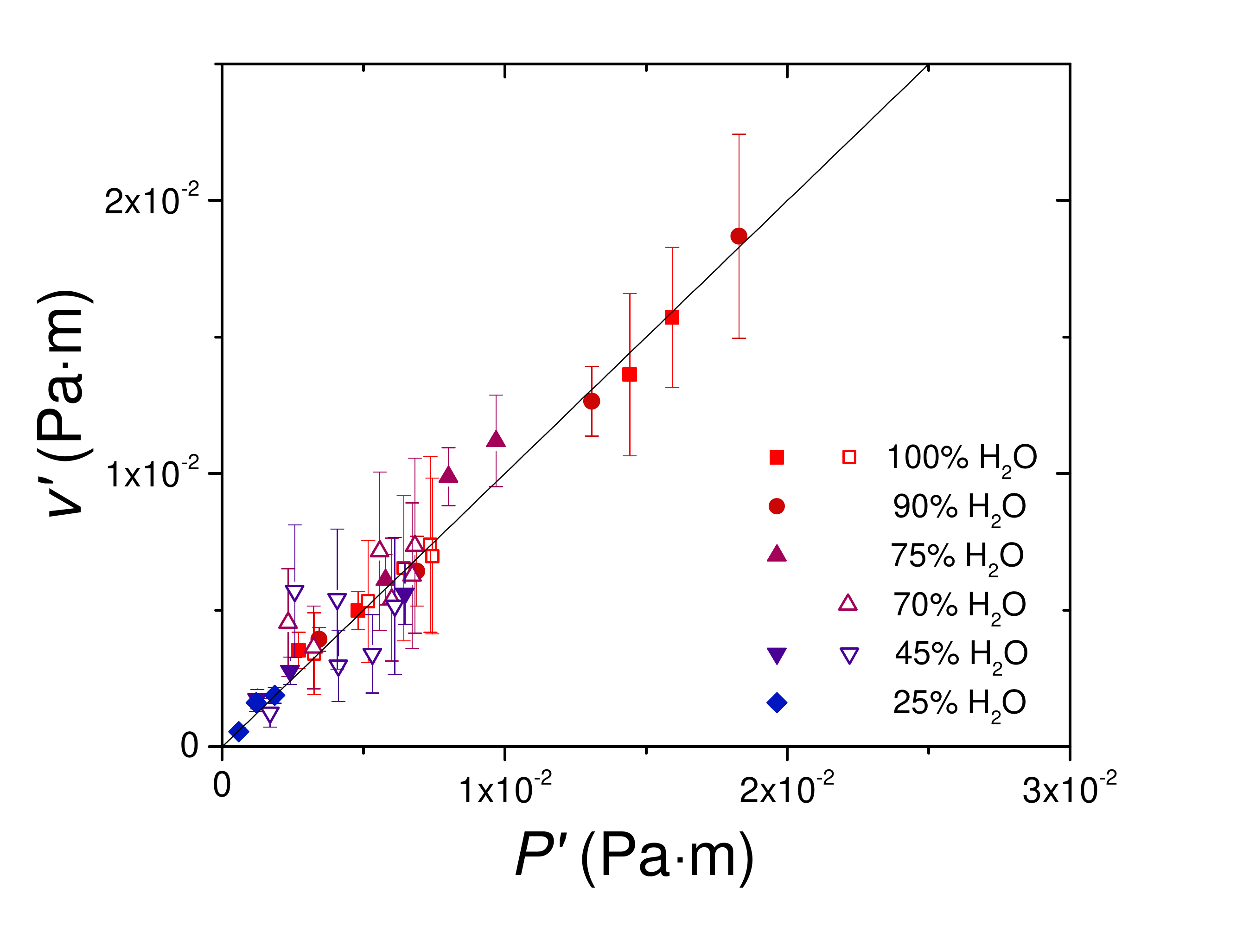}}
	\hfill
	\subfloat[\label{Gamma}]{\includegraphics[width = 0.55\linewidth]{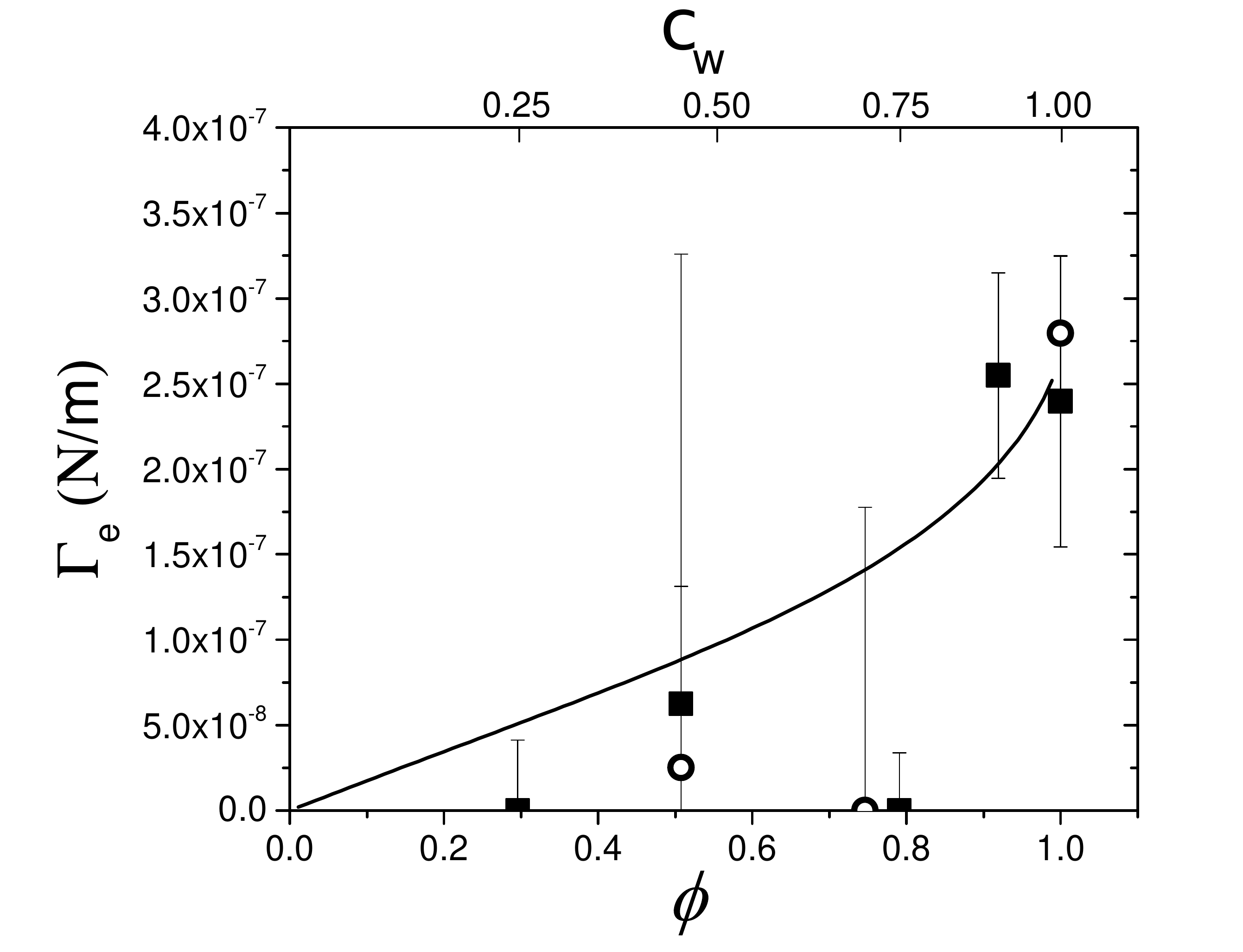}}
	\caption{(a) Mastercurve of the data of Fig. \ref{v_forcing}, obtained by using the scale variable introduced in the text. (b) Experimental values of $\Gammae$ used to rescale the data in panel (a) as a function of the water volume fraction $\phi$ (bottom axis) and the correspondent water mass fraction $c_w$ (top axis) in the drop. The black line is the theoretical prediction for $\Gammae$ (Eq. \ref{gamma-mf}), discussed in the text. In both panels, solid (open) points refer to measurements with fluorescein concentration $\num{2e-3}$ wt/wt ($10^{-3}$ wt/wt).}
	\label{masterc}	
\end{figure}

Albeit simple, Eq. \ref{velocity} allows all the main features of the experimental data of Fig. \ref{v_forcing} to be rationalized: the velocity of the radial deformation at the onset of the instability varies linearly on the centripetal forcing $\Delta P_\omega$, with a prefactor that does not depend on the specific parameters of the drop fluid, namely its viscosity and water concentration, $c_w$. Furthermore, Eq. \ref{velocity} contains an offset proportional to $\Gammae$, which for miscible fluids is expected to depend on the concentration gradient according to Eq. \ref{Korteweg_Gamma}, and thus to be more significant at the highest $c_w$. This explains why the data for $c_w\geq 0.9$ in Fig. \ref{v_forcing} are not compatible with a line passing through the origin.
Figure \ref{mastercurve} shows a mastercurve obtained by fitting each dataset at a given water concentration with Eq. \ref{velocity} and then rescaling the data using the parameters resulting from the best fit, by defining the scaled variables $v'=\left(v\frac{r_0}{l}+\frac{\Gammae}{2\eta_e}\right)\frac{2\eta_e}{\alpha}$ and $P'= r_0 \Delta P_\omega$. In this representation, the data should fall on the straight line $v'=P'$. Figure \ref{masterc} shows that within experimental error this is indeed the case, and that the results do not depend on the fluorescein concentration.

The values of $\Gammae$ used to rescale the data in Fig. \ref{mastercurve} are shown in Fig. \ref{Gamma} as a function of the volume fraction $\phi$ and mass fraction $c_w$ of water in the drop, where solid and open symbols refer to two concentrations of fluorescein. It is worth emphasizing the ultra-low value of the EIT, which attains at most $ 250 \pm 50 \si{\nano\newton/\meter}$ for pure water drops in pure glycerol. This value is much lower than that previously reported in literature \cite{petitjeans_petitjean_1996}. Our findings solve the long-standing controversy stemming from conflicting literature values for the same system \cite{petitjeans_petitjean_1996, legendre_instabilites_2003,paterson_fingering_1985}, as mentioned in section \ref{sec_intro}. In particular, our result is in stark contrast with the value of EIT between water and glycerol reported in \cite{petitjeans_petitjean_1996}, $\Gamma_e = 0.58 \, \si{\milli\newton/\meter}$. Note that this latter value is also in contrast with the experimental observation that drops of pure water spinning in glycerol keep on elongating without reaching a stationary state, even in experiments lasting thousands of seconds. If the EIT was as high as reported in \cite{petitjeans_petitjean_1996}, a drop of water in glycerol would deform towards a stationary state following an exponential relaxation with time constant $\tau \simeq 1.4 \, \si{\second}$, estimated following Ref. \cite{carbonaro_spinning_2019} and assuming a relaxation dynamics similar to that of drops spinning in an immiscible background fluid. This is clearly not the case. By contrast, the ultra-low magnitude of the EIT measured here (a few hundreds of $\si{\nano\newton/\meter}$) is consistent with several works \cite{legendre_instabilites_2003, bischofberger_fingering_2014, lajeunesse_3d_1997} that report a negligible EIT for the same system.

A further support to our findings comes from a review of previous works investigating the Saffman-Taylor instability occurring when water is injected in a Hele-Show cell containing glycerol. Although a Saffman-Taylor instability does occur a few ms after the injection of the less viscous fluid (water) into a Hele-Shaw cell and hence reduce the effects of diffusion, its visualization has systematically suggested that viscous dissipation largely dominates over interfacial effects. In this limit, the wavelength of the instability $\lambda_{ST}$ is only dictated by the gap of the injection cell $b$, and is expected to satisfy
\begin{equation}\label{fingering}
	4b\lesssim\lambda_{ST}\lesssim 5b\,.
\end{equation}
Previous works by Paterson \cite{paterson_fingering_1985} and more recently by Bischofberger and coworkers \cite{bischofberger_fingering_2014} and Lajeunesse and coworkers \cite{lajeunesse_3d_1997} systematically find $\lambda_{ST}$ values in this regime, thus suggesting that the EIT between water and glycerol is too low to be measurable through the Saffman-Taylor instability.
To corroborate this notion, we use Eq. \ref{fingering} to extract a lower bound for the $\Gammae$ values measurable with this technique, which we quantify as the EIT for which the wavelength observed at the onset of the instability equals $4b$. We show in \cite{noauthor_supplemental_nodate} that this lower bound for water/glycerol systems in a typical Hele-Show cell and for accessible injection rates is approximately 0.1 $\si{\milli\newton/\meter}$, while the measurement of much lower values is hampered by diffusion.

By contrast, the analysis of the radial deformation of drops towards dumbbell shapes proposed here allows one to measure EIT values as low as hundreds of $\si{\nano\newton/\meter}$, well beyond the limits of any other standard experimental technique. The strength of the method proposed in this work resides in the fact that the forcing $\alpha \Delta P_\omega$ is very weak, thus allowing relatively low values of the capillary ($Ca$) and the Bond ($Bo$) number to be reached. The first is defined as the ratio between viscous forces and surface tension forces acting across the interface between the fluids: $Ca = \sfrac{\eta v}{\Gamma}$. Considering the viscosity of the background fluid (glycerol), the measured deformation speed $v$ and $\Gamma_e$, the capillary numbers characterizing our experiments are in the range $2\lesssim Ca \lesssim 20$, which are relatively low, taking into account that we probe miscible interfaces and hence $\Gamma_e$ is very small. The second dimensionless number relevant for our experiments is the Bond number, the ratio of the external forces to surface tension forces: $Bo = \sfrac{\alpha\Delta P_{\omega}r_0}{\Gamma_e}$. In our case, $0.4 \lesssim Bo \lesssim 4$, corroborating a scenario where capillary stresses are indeed relevant for the drop deformation. The estimates for $Ca$ and $Bo$ also explain the large uncertainty on the value of $\Gamma_e$ at $\varphi<0.9$ in Fig. \ref{Gamma}: when the EIT decreases to extremely low values, both the capillary and the Bond number increase well above unity, making $\Gamma_e$ barely measurable.

In order to further validate our measurements, we calculate the expected EIT between water and glycerol using a phase field model introduced in \cite{truzzolillo_nonequilibrium_2016}. We briefly recall the main ingredients of the model: by assuming local equilibrium between the two fluids \cite{jinwoo_local_2015,truzzolillo_hydrodynamic_2018,kheniene_linear_2013}, one computes both the enthalpic and the entropic contributions to the Korteweg parameter $\kappa(\tilde{\varphi})$ using lattice theory arguments and assuming for simplicity that the mixture is symmetric, i.e., that the two fluids have the same molecular volume. The two terms are due, respectively, to the variation of the internal energy density $u$ and to the decrease of configurational entropy density $s$ in the region where $|\nabla \tilde{\varphi}|>$ 0. They are obtained by expressing $u$ and $s$ as a function of the local concentration, assuming a concentration gradient across three adjacent lattice layers orthogonal to the $z$ direction, and finally by expanding the local concentration around that of the central layer, up to second order in the spatial derivatives of $\tilde{\varphi}$. Furthermore, in analogy to equilibrium systems, the local concentration profile is modelled by $\tilde{\varphi}=\frac{\phi}{2}+\frac{\phi}{2}\tanh(\frac{z}{\delta})$, with $\phi$ the volume fraction of one kind of molecules, e.g water, in the bulk fluids.
As detailed in Ref. \cite{truzzolillo_nonequilibrium_2016}, the model predicts
\begin{equation}\label{gamma-mf}
\Gamma_{e}=\frac{R Ta^2}{V_m\delta}\left\{\chi_{wg}\frac{\phi^2}{6}+\frac{2}{3}\left[1+\frac{1-\phi}{\phi}\ln(1-\phi)\right]\right\}\,,
\end{equation}
with $R$ the ideal gas constant, $a$ and $V_m$ the diameter and the molar volume of the fluid molecules respectively, and $\delta$ the interface thickness.
The first term on the rhs of Eq. \ref{gamma-mf}, proportional to the $\chi_{wg}$ parameter characterizing the interaction between water and glycerol molecules, quantifies the energy penalty (or gain) due to a local compositional inhomogeneity. The second term is always positive and depends only on $\phi$. It quantifies the entropy loss due to the (transient) gradient of concentration.
We note that the approximations used to derive Eq. \ref{gamma-mf} imply that the concentration gradient at the interface be small, such that the effective interfacial tension be dominated by the first term of the mixing free energy expansion, i.e. the square gradient one.

To estimate $\Gamma_e$ for our water-glycerol mixtures we use Eq. \ref{gamma-mf} and take the average of the molecular diameter and molar volume of water and glycerol: $\langle a\rangle=$0.45 nm and $\langle V_m\rangle$=45.5 ml/mol. We further consider the effect of diffusion by calculating the thickness of the interface as $\delta=\sqrt{2D_{wg} t}$ where $D_{wg}=1.4\cdot10^{-11}$ m$^2$/s is the diffusion coefficient of water in glycerol \cite{derrico_diffusion_2004} and $t\approx 15\, \si{\second}$ is the typical time elapsing between the drop injection and the observation of the instability. The result is displayed in Fig. \ref{Gamma} (line), showing that our data are very well captured by the theoretical $\Gamma_e$ obtained via Eq. \ref{gamma-mf}, with no adjustable parameter. The agreement between the data and Eq. \ref{gamma-mf} suggests that the model of Ref. \cite{truzzolillo_nonequilibrium_2016}, albeit very simple, may be reliably used to estimate the EIT. Since -1$<\chi<$1 for most pairs of miscible substances \cite{barton_handbook_1990}, for $\phi\backsimeq$1 the effective interfacial tension between miscible molecular fluids is well approximated by:
\begin{equation}\label{gamma-mf-approx}
\Gamma_{e}\thickapprox\frac{2R Ta^2}{3V_m\delta}\,.
\end{equation}
As an example, for liquids with characteristics similar to water ($a\thickapprox0.1$ nm, $V_m=18$ ml/mol, $D=2.28\cdot 10^{-9}$ m$^2$/s \cite{walderhaug_aqueous_2012} at $T=298.15$ K), Eq. \ref{gamma-mf-approx} predicts a tension ranging from $\Gamma_{e}\thickapprox 9$ mN/m to $\Gamma_{e}\thickapprox\num{13e-6}$ mN/m as the interface thickness increases from a value comparable to the molecular size ($\delta \thickapprox a$ for $t=$0) to 67 $\mu$m after $t=1$ s of interdiffusion. This supports the fact that for fully miscible low-viscosity fluids capillary effects decay very rapidly with time and can be safely neglected in most of the cases. However, this may not be the case when diffusivity is very low, a condition that can be attained in many simple liquids like silicon oils \cite{kuang_miscible_2003}, colloidal and polymer suspensions \cite{truzzolillo_nonequilibrium_2016}, or in geologically relevant fluids such as silicate fluids in the Earth mantle \cite{morra_role_2008, mungall_interfacial_1994}.
Finally, we note that this argument offers also a possible explanation for the absence of deformation for the triethylene glycol drop in Fig. \ref{Carabinieri}, panel B. Indeed, TEG has a miscibility with glycerol similar to that of water, the three liquids having similar Hansen solubility parameters \cite{hansen_hansen_2000}, but it is significantly more viscous than water ($\eta_{TEG} = 49 \,\si{\milli\pascal\second}$). Therefore, one may expect, on the time scale of our experiments, a stronger effect of the EIT as compared to the case of water drops, thus preventing the development of the dumbbell shape.

\section{Conclusions}\label{sec_conclusions}
We have experimentally characterized the time evolution of the shape of miscible drops in SDT experiments. We have shown that a dumbbell shape arises for sufficiently low values of EIT, which depends not only on the density and viscosity contrast between the drop and the background fluids, but also on the molecular structure of the fluids. We have focussed on mixtures of water and glycerol, for which literature data were conflicting \cite{petitjeans_petitjean_1996, legendre_instabilites_2003, paterson_fingering_1985}, and employed the dynamics of the drop shape as a tool to measure the EIT. By means of a simple model which takes into account the normal stress on the drop surface, the shear stress opposing the deformation and a Laplace-like term containing an effective interfacial tension, we obtained an EIT of $250 \pm 50 \si{\nano\newton/\meter}$ for water in contact with pure glycerol, decreasing rapidly below the resolution limit of the method as the amount of glycerol in the drops increases above 10\% . This result is in excellent agreement with an estimate of the order of magnitude for the EIT for the same system obtained from a phase field model \cite{truzzolillo_nonequilibrium_2016}, while it is orders of magnitude lower than the experimental limit of all conventional techniques for measuring interfacial tensions. Therefore, besides shedding light on the controversy stemming from conflicting literature data on the EIT between water and glycerol, our work demonstrates a new method to measure extremely low interfacial tensions and in particular the EIT, paving the way for a thorough understanding of Korteweg stresses and capillary phenomena in miscible fluids.

\section{Acknowledgements}
We gratefully acknowledge support from the Centre national d'\'{e}tudes spatiales (CNES). We thank Lucio Isa for fruitful discussions.


\begin{thebibliography}{11}
\expandafter\ifx\csname natexlab\endcsname\relax\def\natexlab#1{#1}\fi
\expandafter\ifx\csname bibnamefont\endcsname\relax
  \def\bibnamefont#1{#1}\fi
\expandafter\ifx\csname bibfnamefont\endcsname\relax
  \def\bibfnamefont#1{#1}\fi
\expandafter\ifx\csname citenamefont\endcsname\relax
  \def\citenamefont#1{#1}\fi
\expandafter\ifx\csname url\endcsname\relax
  \def\url#1{\texttt{#1}}\fi
\expandafter\ifx\csname urlprefix\endcsname\relax\def\urlprefix{URL }\fi
\providecommand{\bibinfo}[2]{#2}
\providecommand{\eprint}[2][]{\url{#2}}

\bibitem[{\citenamefont{{M. R. Spiegel} and
  Stephens}(2008)}]{m_r_spiegel_statistics_2008}
\bibinfo{author}{\bibnamefont{{M. R. Spiegel}}} \bibnamefont{and}
  \bibinfo{author}{\bibfnamefont{L.~J.} \bibnamefont{Stephens}},
  \emph{\bibinfo{title}{Statistics}} (\bibinfo{publisher}{McGraw-Hill},
  \bibinfo{year}{2008}), \bibinfo{edition}{schaum} ed.

\bibitem[{\citenamefont{Press et~al.}(1996)\citenamefont{Press, Teukolsky,
  Vetterling, and Flannery}}]{press_numerical_1996}
\bibinfo{author}{\bibfnamefont{W.~H.} \bibnamefont{Press}},
  \bibinfo{author}{\bibfnamefont{S.~A.} \bibnamefont{Teukolsky}},
  \bibinfo{author}{\bibfnamefont{W.~T.} \bibnamefont{Vetterling}},
  \bibnamefont{and} \bibinfo{author}{\bibfnamefont{B.~P.}
  \bibnamefont{Flannery}}, \emph{\bibinfo{title}{Numerical recipes in fortran:
  the art of scientific computing}} (\bibinfo{publisher}{Cambridge University
  Press}, \bibinfo{year}{1996}), ISBN \bibinfo{isbn}{978-0-521-43064-7
  978-0-521-57439-6}.

\bibitem[{\citenamefont{Currie and Van~Nieuwkoop}(1982)}]{currie_buoyancy_1982}
\bibinfo{author}{\bibfnamefont{P.}~\bibnamefont{Currie}} \bibnamefont{and}
  \bibinfo{author}{\bibfnamefont{J.}~\bibnamefont{Van~Nieuwkoop}},
  \bibinfo{journal}{Journal of Colloid and Interface Science}
  \textbf{\bibinfo{volume}{87}}, \bibinfo{pages}{301} (\bibinfo{year}{1982}),
  ISSN \bibinfo{issn}{00219797},
  \urlprefix\url{https://linkinghub.elsevier.com/retrieve/pii/0021979782903289}.

\bibitem[{\citenamefont{Manning and Scriven}(1977)}]{manning_interfacial_1977}
\bibinfo{author}{\bibfnamefont{C.~D.} \bibnamefont{Manning}} \bibnamefont{and}
  \bibinfo{author}{\bibfnamefont{L.~E.} \bibnamefont{Scriven}},
  \bibinfo{journal}{Review of Scientific Instruments}
  \textbf{\bibinfo{volume}{48}}, \bibinfo{pages}{1699} (\bibinfo{year}{1977}),
  ISSN \bibinfo{issn}{0034-6748, 1089-7623},
  \urlprefix\url{http://aip.scitation.org/doi/10.1063/1.1134934}.

\bibitem[{\citenamefont{Zoltowski et~al.}(2007)\citenamefont{Zoltowski,
  Chekanov, Masere, Pojman, and Volpert}}]{zoltowski_evidence_2007}
\bibinfo{author}{\bibfnamefont{B.}~\bibnamefont{Zoltowski}},
  \bibinfo{author}{\bibfnamefont{Y.}~\bibnamefont{Chekanov}},
  \bibinfo{author}{\bibfnamefont{J.}~\bibnamefont{Masere}},
  \bibinfo{author}{\bibfnamefont{J.~A.} \bibnamefont{Pojman}},
  \bibnamefont{and} \bibinfo{author}{\bibfnamefont{V.}~\bibnamefont{Volpert}},
  \bibinfo{journal}{Langmuir} \textbf{\bibinfo{volume}{23}},
  \bibinfo{pages}{5522} (\bibinfo{year}{2007}), ISSN \bibinfo{issn}{0743-7463,
  1520-5827}, \urlprefix\url{https://pubs.acs.org/doi/10.1021/la063382g}.

\bibitem[{\citenamefont{Quarteroni}(2017)}]{quarteroni_numerical_2017}
\bibinfo{author}{\bibfnamefont{A.}~\bibnamefont{Quarteroni}},
  \emph{\bibinfo{title}{Numerical {Models} for {Differential} {Problems}}},
  vol.~\bibinfo{volume}{16} of \emph{\bibinfo{series}{{MS}\&{A}}}
  (\bibinfo{publisher}{Springer International Publishing},
  \bibinfo{address}{Cham}, \bibinfo{year}{2017}), ISBN
  \bibinfo{isbn}{978-3-319-49315-2 978-3-319-49316-9},
  \urlprefix\url{http://link.springer.com/10.1007/978-3-319-49316-9}.

\bibitem[{\citenamefont{Truzzolillo et~al.}(2014)\citenamefont{Truzzolillo,
  Mora, Dupas, and Cipelletti}}]{truzzolillo_off-equilibrium_2014}
\bibinfo{author}{\bibfnamefont{D.}~\bibnamefont{Truzzolillo}},
  \bibinfo{author}{\bibfnamefont{S.}~\bibnamefont{Mora}},
  \bibinfo{author}{\bibfnamefont{C.}~\bibnamefont{Dupas}}, \bibnamefont{and}
  \bibinfo{author}{\bibfnamefont{L.}~\bibnamefont{Cipelletti}},
  \bibinfo{journal}{Physical Review Letters} \textbf{\bibinfo{volume}{112}}
  (\bibinfo{year}{2014}), ISSN \bibinfo{issn}{0031-9007, 1079-7114},
  \urlprefix\url{https://link.aps.org/doi/10.1103/PhysRevLett.112.128303}.

\bibitem[{\citenamefont{Truzzolillo et~al.}(2016)\citenamefont{Truzzolillo,
  Mora, Dupas, and Cipelletti}}]{truzzolillo_nonequilibrium_2016}
\bibinfo{author}{\bibfnamefont{D.}~\bibnamefont{Truzzolillo}},
  \bibinfo{author}{\bibfnamefont{S.}~\bibnamefont{Mora}},
  \bibinfo{author}{\bibfnamefont{C.}~\bibnamefont{Dupas}}, \bibnamefont{and}
  \bibinfo{author}{\bibfnamefont{L.}~\bibnamefont{Cipelletti}},
  \bibinfo{journal}{Physical Review X} \textbf{\bibinfo{volume}{6}},
  \bibinfo{pages}{041057} (\bibinfo{year}{2016}).

\bibitem[{\citenamefont{Paterson}(1985)}]{paterson_fingering_1985}
\bibinfo{author}{\bibfnamefont{L.}~\bibnamefont{Paterson}},
  \bibinfo{journal}{Physics of Fluids} \textbf{\bibinfo{volume}{28}},
  \bibinfo{pages}{26} (\bibinfo{year}{1985}), ISSN \bibinfo{issn}{0031-9171},
  \urlprefix\url{https://aip.scitation.org/doi/10.1063/1.865195}.

\bibitem[{\citenamefont{Bischofberger et~al.}(2014)\citenamefont{Bischofberger,
  Ramachandran, and Nagel}}]{bischofberger_fingering_2014}
\bibinfo{author}{\bibfnamefont{I.}~\bibnamefont{Bischofberger}},
  \bibinfo{author}{\bibfnamefont{R.}~\bibnamefont{Ramachandran}},
  \bibnamefont{and} \bibinfo{author}{\bibfnamefont{S.~R.} \bibnamefont{Nagel}},
  \bibinfo{journal}{Nature Communications} \textbf{\bibinfo{volume}{5}}
  (\bibinfo{year}{2014}), ISSN \bibinfo{issn}{2041-1723},
  \urlprefix\url{http://www.nature.com/articles/ncomms6265}.

\bibitem[{\citenamefont{Miranda and Widom}(1998)}]{miranda_radial_1998}
\bibinfo{author}{\bibfnamefont{J.}~\bibnamefont{Miranda}} \bibnamefont{and}
  \bibinfo{author}{\bibfnamefont{M.}~\bibnamefont{Widom}},
  \bibinfo{journal}{Physica D: Nonlinear Phenomena}
  \textbf{\bibinfo{volume}{120}}, \bibinfo{pages}{315} (\bibinfo{year}{1998}),
  ISSN \bibinfo{issn}{01672789},
  \urlprefix\url{https://linkinghub.elsevier.com/retrieve/pii/S0167278998000979}.

\end{thebibliography}


\begin{thebibliography}{45}
\expandafter\ifx\csname natexlab\endcsname\relax\def\natexlab#1{#1}\fi
\expandafter\ifx\csname bibnamefont\endcsname\relax
  \def\bibnamefont#1{#1}\fi
\expandafter\ifx\csname bibfnamefont\endcsname\relax
  \def\bibfnamefont#1{#1}\fi
\expandafter\ifx\csname citenamefont\endcsname\relax
  \def\citenamefont#1{#1}\fi
\expandafter\ifx\csname url\endcsname\relax
  \def\url#1{\texttt{#1}}\fi
\expandafter\ifx\csname urlprefix\endcsname\relax\def\urlprefix{URL }\fi
\providecommand{\bibinfo}[2]{#2}
\providecommand{\eprint}[2][]{\url{#2}}

\bibitem[{\citenamefont{Rowlinson and Widom}(2002)}]{rowlinson_molecular_2002}
\bibinfo{author}{\bibfnamefont{J.~S.} \bibnamefont{Rowlinson}}
  \bibnamefont{and} \bibinfo{author}{\bibfnamefont{B.}~\bibnamefont{Widom}},
  \emph{\bibinfo{title}{Molecular theory of capillarity}}
  (\bibinfo{publisher}{Dover Publications}, \bibinfo{address}{Mineola, N.Y},
  \bibinfo{year}{2002}), ISBN \bibinfo{isbn}{978-0-486-42544-3}.

\bibitem[{\citenamefont{Korteweg}(1901)}]{korteweg_kortewegpdf_1901}
\bibinfo{author}{\bibfnamefont{D.~J.} \bibnamefont{Korteweg}},
  \bibinfo{journal}{Archives N�erlandaises} \textbf{\bibinfo{volume}{2}},
  \bibinfo{pages}{1} (\bibinfo{year}{1901}).

\bibitem[{\citenamefont{Cahn and Hilliard}(1958)}]{cahn_free_1958}
\bibinfo{author}{\bibfnamefont{J.~W.} \bibnamefont{Cahn}} \bibnamefont{and}
  \bibinfo{author}{\bibfnamefont{J.~E.} \bibnamefont{Hilliard}},
  \bibinfo{journal}{The Journal of Chemical Physics}
  \textbf{\bibinfo{volume}{28}}, \bibinfo{pages}{11} (\bibinfo{year}{1958}).

\bibitem[{\citenamefont{Truzzolillo and
  Cipelletti}(2017)}]{truzzolillo_off-equilibrium_2017}
\bibinfo{author}{\bibfnamefont{D.}~\bibnamefont{Truzzolillo}} \bibnamefont{and}
  \bibinfo{author}{\bibfnamefont{L.}~\bibnamefont{Cipelletti}},
  \bibinfo{journal}{Soft Matter} \textbf{\bibinfo{volume}{13}},
  \bibinfo{pages}{13} (\bibinfo{year}{2017}), ISSN \bibinfo{issn}{1744-683X,
  1744-6848}, \urlprefix\url{http://xlink.rsc.org/?DOI=C6SM01026A}.

\bibitem[{\citenamefont{Truzzolillo et~al.}(2016)\citenamefont{Truzzolillo,
  Mora, Dupas, and Cipelletti}}]{truzzolillo_nonequilibrium_2016}
\bibinfo{author}{\bibfnamefont{D.}~\bibnamefont{Truzzolillo}},
  \bibinfo{author}{\bibfnamefont{S.}~\bibnamefont{Mora}},
  \bibinfo{author}{\bibfnamefont{C.}~\bibnamefont{Dupas}}, \bibnamefont{and}
  \bibinfo{author}{\bibfnamefont{L.}~\bibnamefont{Cipelletti}},
  \bibinfo{journal}{Physical Review X} \textbf{\bibinfo{volume}{6}},
  \bibinfo{pages}{041057} (\bibinfo{year}{2016}).

\bibitem[{\citenamefont{Cicuta et~al.}(2001)\citenamefont{Cicuta, Vailati, and
  Giglio}}]{cicuta_capillary--bulk_2001}
\bibinfo{author}{\bibfnamefont{P.}~\bibnamefont{Cicuta}},
  \bibinfo{author}{\bibfnamefont{A.}~\bibnamefont{Vailati}}, \bibnamefont{and}
  \bibinfo{author}{\bibfnamefont{M.}~\bibnamefont{Giglio}},
  \bibinfo{journal}{Applied Optics} \textbf{\bibinfo{volume}{40}},
  \bibinfo{pages}{4140} (\bibinfo{year}{2001}), ISSN \bibinfo{issn}{0003-6935,
  1539-4522},
  \urlprefix\url{https://www.osapublishing.org/abstract.cfm?URI=ao-40-24-4140}.

\bibitem[{\citenamefont{Vorobev et~al.}(2020)\citenamefont{Vorobev, Zagvozkin,
  and Lyubimova}}]{vorobev_shapes_2020}
\bibinfo{author}{\bibfnamefont{A.}~\bibnamefont{Vorobev}},
  \bibinfo{author}{\bibfnamefont{T.}~\bibnamefont{Zagvozkin}},
  \bibnamefont{and}
  \bibinfo{author}{\bibfnamefont{T.}~\bibnamefont{Lyubimova}},
  \bibinfo{journal}{Physics of Fluids} \textbf{\bibinfo{volume}{32}},
  \bibinfo{pages}{012112} (\bibinfo{year}{2020}), ISSN
  \bibinfo{issn}{1070-6631, 1089-7666},
  \urlprefix\url{http://aip.scitation.org/doi/10.1063/1.5141334}.

\bibitem[{\citenamefont{Antanovskii}(1996)}]{antanovskii_microscale_1996}
\bibinfo{author}{\bibfnamefont{L.~K.} \bibnamefont{Antanovskii}},
  \bibinfo{journal}{Physical Review E} \textbf{\bibinfo{volume}{54}},
  \bibinfo{pages}{6285} (\bibinfo{year}{1996}), ISSN \bibinfo{issn}{1063-651X,
  1095-3787},
  \urlprefix\url{https://link.aps.org/doi/10.1103/PhysRevE.54.6285}.

\bibitem[{\citenamefont{Bier}(2015)}]{bier_nonequilibrium_2015}
\bibinfo{author}{\bibfnamefont{M.}~\bibnamefont{Bier}},
  \bibinfo{journal}{Physical Review E} \textbf{\bibinfo{volume}{92}},
  \bibinfo{pages}{042128} (\bibinfo{year}{2015}), ISSN \bibinfo{issn}{1539-3755, 1550-2376},
  \urlprefix\url{https://link.aps.org/doi/10.1103/PhysRevE.92.042128}.

\bibitem[{\citenamefont{Vorobev and Boghi}(2016)}]{vorobev_phase-field_2016}
\bibinfo{author}{\bibfnamefont{A.}~\bibnamefont{Vorobev}} \bibnamefont{and}
  \bibinfo{author}{\bibfnamefont{A.}~\bibnamefont{Boghi}},
  \bibinfo{journal}{Journal of Colloid and Interface Science}
  \textbf{\bibinfo{volume}{482}}, \bibinfo{pages}{193} (\bibinfo{year}{2016}),
  ISSN \bibinfo{issn}{00219797},
  \urlprefix\url{https://linkinghub.elsevier.com/retrieve/pii/S0021979716305471}.

\bibitem[{\citenamefont{Chen and Meiburg}(1996)}]{chen_miscible_1996}
\bibinfo{author}{\bibfnamefont{C.-Y.} \bibnamefont{Chen}} \bibnamefont{and}
  \bibinfo{author}{\bibfnamefont{E.}~\bibnamefont{Meiburg}},
  \bibinfo{journal}{Journal of Fluid Mechanics} \textbf{\bibinfo{volume}{326}},
  \bibinfo{pages}{57} (\bibinfo{year}{1996}), ISSN \bibinfo{issn}{0022-1120,
  1469-7645},
  \urlprefix\url{https://www.cambridge.org/core/product/identifier/S0022112096008245/type/journal_article}.

\bibitem[{\citenamefont{Chen and Meiburg}(2002)}]{chen_miscible_2002}
\bibinfo{author}{\bibfnamefont{C.-Y.} \bibnamefont{Chen}} \bibnamefont{and}
  \bibinfo{author}{\bibfnamefont{E.}~\bibnamefont{Meiburg}},
  \bibinfo{journal}{Physics of Fluids} \textbf{\bibinfo{volume}{14}},
  \bibinfo{pages}{2052} (\bibinfo{year}{2002}), ISSN \bibinfo{issn}{10706631},
  \urlprefix\url{http://scitation.aip.org/content/aip/journal/pof2/14/7/10.1063/1.1481507}.

\bibitem[{\citenamefont{Shevtsova et~al.}(2016)\citenamefont{Shevtsova,
  Gaponenko, Yasnou, Mialdun, and Nepomnyashchy}}]{shevtsova_two-scale_2016}
\bibinfo{author}{\bibfnamefont{V.}~\bibnamefont{Shevtsova}},
  \bibinfo{author}{\bibfnamefont{Y.~A.} \bibnamefont{Gaponenko}},
  \bibinfo{author}{\bibfnamefont{V.}~\bibnamefont{Yasnou}},
  \bibinfo{author}{\bibfnamefont{A.}~\bibnamefont{Mialdun}}, \bibnamefont{and}
  \bibinfo{author}{\bibfnamefont{A.}~\bibnamefont{Nepomnyashchy}},
  \bibinfo{journal}{J. Fluid Mech.} \textbf{\bibinfo{volume}{795}},
  \bibinfo{pages}{409} (\bibinfo{year}{2016}), ISSN \bibinfo{issn}{0022-1120,
  1469-7645},
  \urlprefix\url{https://www.cambridge.org/core/product/identifier/S0022112016002226/type/journal_article}.

\bibitem[{\citenamefont{May and Maher}(1991)}]{may_capillary-wave_1991}
\bibinfo{author}{\bibfnamefont{S.~E.}~\bibnamefont{May}} \bibnamefont{and}
  \bibinfo{author}{\bibfnamefont{J.~V.}~\bibnamefont{Maher}},
  \bibinfo{journal}{Physical Review Letters} \textbf{\bibinfo{volume}{67}},
  \bibinfo{pages}{2013} (\bibinfo{year}{1991}), ISSN \bibinfo{issn}{0031-9007},
  \urlprefix\url{https://link.aps.org/doi/10.1103/PhysRevLett.67.2013}.

\bibitem[{\citenamefont{Pojman et~al.}(2006)\citenamefont{Pojman, Whitmore,
  Turco~Liveri, Lombardo, Marszalek, Parker, and
  Zoltowski}}]{pojman_evidence_2006}
\bibinfo{author}{\bibfnamefont{J.~A.} \bibnamefont{Pojman}},
  \bibinfo{author}{\bibfnamefont{C.}~\bibnamefont{Whitmore}},
  \bibinfo{author}{\bibfnamefont{M.~L.} \bibnamefont{Turco~Liveri}},
  \bibinfo{author}{\bibfnamefont{R.}~\bibnamefont{Lombardo}},
  \bibinfo{author}{\bibfnamefont{J.}~\bibnamefont{Marszalek}},
  \bibinfo{author}{\bibfnamefont{R.}~\bibnamefont{Parker}}, \bibnamefont{and}
  \bibinfo{author}{\bibfnamefont{B.}~\bibnamefont{Zoltowski}},
  \bibinfo{journal}{Langmuir} \textbf{\bibinfo{volume}{22}},
  \bibinfo{pages}{2569} (\bibinfo{year}{2006}), ISSN \bibinfo{issn}{0743-7463,
  1520-5827}, \urlprefix\url{http://pubs.acs.org/doi/abs/10.1021/la052111n}.

\bibitem[{\citenamefont{Carbonaro et~al.}(2019)\citenamefont{Carbonaro,
  Cipelletti, and Truzzolillo}}]{carbonaro_spinning_2019}
\bibinfo{author}{\bibfnamefont{A.}~\bibnamefont{Carbonaro}},
  \bibinfo{author}{\bibfnamefont{L.}~\bibnamefont{Cipelletti}},
  \bibnamefont{and}
  \bibinfo{author}{\bibfnamefont{D.}~\bibnamefont{Truzzolillo}},
  \bibinfo{journal}{Langmuir} \textbf{\bibinfo{volume}{35}},
  \bibinfo{pages}{11330} (\bibinfo{year}{2019}), ISSN \bibinfo{issn}{0743-7463,
  1520-5827},
  \urlprefix\url{http://pubs.acs.org/doi/10.1021/acs.langmuir.9b02091}.

\bibitem[{\citenamefont{Gowda.~V et~al.}(2019)\citenamefont{Gowda.~V, Brouzet,
  Lefranc, S�derberg, and Lundell}}]{gowda_v_effective_2019}
\bibinfo{author}{\bibfnamefont{K.}~\bibnamefont{Gowda.~V}},
  \bibinfo{author}{\bibfnamefont{C.}~\bibnamefont{Brouzet}},
  \bibinfo{author}{\bibfnamefont{T.}~\bibnamefont{Lefranc}},
  \bibinfo{author}{\bibfnamefont{L.~D.} \bibnamefont{S�derberg}},
  \bibnamefont{and} \bibinfo{author}{\bibfnamefont{F.}~\bibnamefont{Lundell}},
  \bibinfo{journal}{J. Fluid Mech.} \textbf{\bibinfo{volume}{876}},
  \bibinfo{pages}{1052} (\bibinfo{year}{2019}), ISSN \bibinfo{issn}{0022-1120,
  1469-7645},
  \urlprefix\url{https://www.cambridge.org/core/product/identifier/S0022112019005664/type/journal_article}.

\bibitem[{\citenamefont{Bamberger et~al.}(1984)\citenamefont{Bamberger, Seaman,
  Sharp, and Brooks}}]{bamberger_effects_1984}
\bibinfo{author}{\bibfnamefont{S.}~\bibnamefont{Bamberger}},
  \bibinfo{author}{\bibfnamefont{G.~V.} \bibnamefont{Seaman}},
  \bibinfo{author}{\bibfnamefont{K.}~\bibnamefont{Sharp}}, \bibnamefont{and}
  \bibinfo{author}{\bibfnamefont{D.~E.} \bibnamefont{Brooks}},
  \bibinfo{journal}{Journal of Colloid and Interface Science}
  \textbf{\bibinfo{volume}{99}}, \bibinfo{pages}{194} (\bibinfo{year}{1984}),
  ISSN \bibinfo{issn}{00219797},
  \urlprefix\url{https://linkinghub.elsevier.com/retrieve/pii/0021979784901000}.

\bibitem[{\citenamefont{Liu et~al.}(2012)\citenamefont{Liu, Lipowsky, and
  Dimova}}]{liu_concentration_2012}
\bibinfo{author}{\bibfnamefont{Y.}~\bibnamefont{Liu}},
  \bibinfo{author}{\bibfnamefont{R.}~\bibnamefont{Lipowsky}}, \bibnamefont{and}
  \bibinfo{author}{\bibfnamefont{R.}~\bibnamefont{Dimova}},
  \bibinfo{journal}{Langmuir} \textbf{\bibinfo{volume}{28}},
  \bibinfo{pages}{3831} (\bibinfo{year}{2012}), ISSN \bibinfo{issn}{0743-7463,
  1520-5827}, \urlprefix\url{https://pubs.acs.org/doi/10.1021/la204757z}.

\bibitem[{\citenamefont{Vonnegut}(1942)}]{vonnegut_rotating_1942}
\bibinfo{author}{\bibfnamefont{B.}~\bibnamefont{Vonnegut}},
  \bibinfo{journal}{Review of Scientific Instruments}
  \textbf{\bibinfo{volume}{13}}, \bibinfo{pages}{6} (\bibinfo{year}{1942}),
  ISSN \bibinfo{issn}{0034-6748, 1089-7623},
  \urlprefix\url{http://aip.scitation.org/doi/10.1063/1.1769937}.

\bibitem[{\citenamefont{Zoltowski et~al.}(2007)\citenamefont{Zoltowski,
  Chekanov, Masere, Pojman, and Volpert}}]{zoltowski_evidence_2007}
\bibinfo{author}{\bibfnamefont{B.}~\bibnamefont{Zoltowski}},
  \bibinfo{author}{\bibfnamefont{Y.}~\bibnamefont{Chekanov}},
  \bibinfo{author}{\bibfnamefont{J.}~\bibnamefont{Masere}},
  \bibinfo{author}{\bibfnamefont{J.~A.} \bibnamefont{Pojman}},
  \bibnamefont{and} \bibinfo{author}{\bibfnamefont{V.}~\bibnamefont{Volpert}},
  \bibinfo{journal}{Langmuir} \textbf{\bibinfo{volume}{23}},
  \bibinfo{pages}{5522} (\bibinfo{year}{2007}), ISSN \bibinfo{issn}{0743-7463,
  1520-5827}, \urlprefix\url{https://pubs.acs.org/doi/10.1021/la063382g}.

\bibitem[{\citenamefont{Petitjeans}(1996)}]{petitjeans_petitjean_1996}
\bibinfo{author}{\bibfnamefont{P.}~\bibnamefont{Petitjeans}},
  \bibinfo{journal}{C.R. Acad Sci. Paris} pp. \bibinfo{pages}{673--679}
  (\bibinfo{year}{1996}).

\bibitem[{\citenamefont{Legendre et~al.}(2003)\citenamefont{Legendre,
  Petitjeans, and Kurowski}}]{legendre_instabilites_2003}
\bibinfo{author}{\bibfnamefont{M.}~\bibnamefont{Legendre}},
  \bibinfo{author}{\bibfnamefont{P.}~\bibnamefont{Petitjeans}},
  \bibnamefont{and} \bibinfo{author}{\bibfnamefont{P.}~\bibnamefont{Kurowski}},
  \bibinfo{journal}{Comptes Rendus M�canique} \textbf{\bibinfo{volume}{331}},
  \bibinfo{pages}{617} (\bibinfo{year}{2003}), ISSN \bibinfo{issn}{16310721},
  \urlprefix\url{https://linkinghub.elsevier.com/retrieve/pii/S163107210300127X}.

\bibitem[{\citenamefont{Manning and Scriven}(1977)}]{manning_interfacial_1977}
\bibinfo{author}{\bibfnamefont{C.~D.} \bibnamefont{Manning}} \bibnamefont{and}
  \bibinfo{author}{\bibfnamefont{L.~E.} \bibnamefont{Scriven}},
  \bibinfo{journal}{Review of Scientific Instruments}
  \textbf{\bibinfo{volume}{48}}, \bibinfo{pages}{1699} (\bibinfo{year}{1977}),
  ISSN \bibinfo{issn}{0034-6748, 1089-7623},
  \urlprefix\url{http://aip.scitation.org/doi/10.1063/1.1134934}.

\bibitem[{noa(1963)}]{noauthor_physical_1963}
\emph{\bibinfo{title}{Physical properties of glycerine and its solutions}}
  (\bibinfo{publisher}{Glycerine Producers' Association}, \bibinfo{address}{New
  York}, \bibinfo{year}{1963}).

\bibitem[{\citenamefont{D'Errico
  et~al.}(2004{\natexlab{a}})\citenamefont{D'Errico, Ortona, Capuano, and
  Vitagliano}}]{derrico_diffusion_2004-1}
\bibinfo{author}{\bibfnamefont{G.}~\bibnamefont{D'Errico}},
  \bibinfo{author}{\bibfnamefont{O.}~\bibnamefont{Ortona}},
  \bibinfo{author}{\bibfnamefont{F.}~\bibnamefont{Capuano}}, \bibnamefont{and}
  \bibinfo{author}{\bibfnamefont{V.}~\bibnamefont{Vitagliano}},
  \bibinfo{journal}{Journal of Chemical \& Engineering Data}
  \textbf{\bibinfo{volume}{49}}, \bibinfo{pages}{1665}
  (\bibinfo{year}{2004}{\natexlab{a}}), ISSN \bibinfo{issn}{0021-9568,
  1520-5134}, \urlprefix\url{https://pubs.acs.org/doi/10.1021/je049917u}.

\bibitem[{\citenamefont{Galambos and
  Forster}(1998)}]{harrison_micro-fluidic_1998}
\bibinfo{author}{\bibfnamefont{P.}~\bibnamefont{Galambos}} \bibnamefont{and}
  \bibinfo{author}{\bibfnamefont{F.~K.} \bibnamefont{Forster}}, in
  \emph{\bibinfo{booktitle}{Micro {Total} {Analysis} {Systems} �98}}, edited
  by \bibinfo{editor}{\bibfnamefont{D.~J.} \bibnamefont{Harrison}}
  \bibnamefont{and} \bibinfo{editor}{\bibfnamefont{A.}~\bibnamefont{van~den
  Berg}} (\bibinfo{publisher}{Springer Netherlands},
  \bibinfo{address}{Dordrecht}, \bibinfo{year}{1998}), pp.
  \bibinfo{pages}{189--192}, ISBN \bibinfo{isbn}{978-94-010-6225-1
  978-94-011-5286-0},
  \urlprefix\url{http://link.springer.com/10.1007/978-94-011-5286-0_46}.

\bibitem[{\citenamefont{Lister and Stone}(1996)}]{lister_time-dependent_1996}
\bibinfo{author}{\bibfnamefont{J.}~\bibnamefont{Lister}} \bibnamefont{and}
  \bibinfo{author}{\bibfnamefont{H.}~\bibnamefont{Stone}},
  \bibinfo{journal}{Journal of Fluid Mechanics} \textbf{\bibinfo{volume}{317}},
  \bibinfo{pages}{275} (\bibinfo{year}{1996}).

\bibitem[{\citenamefont{{M. R. Spiegel} and
  Stephens}(2008)}]{m_r_spiegel_statistics_2008}
\bibinfo{author}{\bibnamefont{{M. R. Spiegel}}} \bibnamefont{and}
  \bibinfo{author}{\bibfnamefont{L.~J.} \bibnamefont{Stephens}},
  \emph{\bibinfo{title}{Statistics}} (\bibinfo{publisher}{McGraw-Hill},
  \bibinfo{year}{2008}), \bibinfo{edition}{schaum} ed.

\bibitem[{\citenamefont{Press et~al.}(1996)\citenamefont{Press, Teukolsky,
  Vetterling, and Flannery}}]{press_numerical_1996}
\bibinfo{author}{\bibfnamefont{W.~H.} \bibnamefont{Press}},
  \bibinfo{author}{\bibfnamefont{S.~A.} \bibnamefont{Teukolsky}},
  \bibinfo{author}{\bibfnamefont{W.~T.} \bibnamefont{Vetterling}},
  \bibnamefont{and} \bibinfo{author}{\bibfnamefont{B.~P.}
  \bibnamefont{Flannery}}, \emph{\bibinfo{title}{Numerical recipes in fortran:
  the art of scientific computing}} (\bibinfo{publisher}{Cambridge University
  Press}, \bibinfo{year}{1996}), ISBN \bibinfo{isbn}{978-0-521-43064-7
  978-0-521-57439-6}.

\bibitem[{noa()}]{noauthor_supplemental_nodate}
\emph{\bibinfo{title}{Supplemental material: We provide details on i) the
  statistical significance of the {{$c_w$}}-dependence of the deformation
  velocity, ii) the simulations of the recirculating flow on top of spinning
  drops and the dependence of the flow on {{$\Delta P_\omega$}} and iii) the
  limitations of the direct visualization of the radial Saffman-Taylor
  instability as a method to measure the effective interfacial tension in
  water-glycerol systems.}}

\bibitem[{\citenamefont{Currie and Van~Nieuwkoop}(1982)}]{currie_buoyancy_1982}
\bibinfo{author}{\bibfnamefont{P.}~\bibnamefont{Currie}} \bibnamefont{and}
  \bibinfo{author}{\bibfnamefont{J.}~\bibnamefont{Van~Nieuwkoop}},
  \bibinfo{journal}{Journal of Colloid and Interface Science}
  \textbf{\bibinfo{volume}{87}}, \bibinfo{pages}{301} (\bibinfo{year}{1982}),
  ISSN \bibinfo{issn}{00219797},
  \urlprefix\url{https://linkinghub.elsevier.com/retrieve/pii/0021979782903289}.

\bibitem[{\citenamefont{Paterson}(1985)}]{paterson_fingering_1985}
\bibinfo{author}{\bibfnamefont{L.}~\bibnamefont{Paterson}},
  \bibinfo{journal}{Physics of Fluids} \textbf{\bibinfo{volume}{28}},
  \bibinfo{pages}{26} (\bibinfo{year}{1985}), ISSN \bibinfo{issn}{0031-9171},
  \urlprefix\url{https://aip.scitation.org/doi/10.1063/1.865195}.

\bibitem[{\citenamefont{Bischofberger et~al.}(2014)\citenamefont{Bischofberger,
  Ramachandran, and Nagel}}]{bischofberger_fingering_2014}
\bibinfo{author}{\bibfnamefont{I.}~\bibnamefont{Bischofberger}},
  \bibinfo{author}{\bibfnamefont{R.}~\bibnamefont{Ramachandran}},
  \bibnamefont{and} \bibinfo{author}{\bibfnamefont{S.~R.} \bibnamefont{Nagel}},
  \bibinfo{journal}{Nature Communications} \textbf{\bibinfo{volume}{5}}
  (\bibinfo{year}{2014}), ISSN \bibinfo{issn}{2041-1723},
  \urlprefix\url{http://www.nature.com/articles/ncomms6265}.

\bibitem[{\citenamefont{Lajeunesse et~al.}(1997)\citenamefont{Lajeunesse,
  Martin, Rakotomalala, and Salin}}]{lajeunesse_3d_1997}
\bibinfo{author}{\bibfnamefont{E.}~\bibnamefont{Lajeunesse}},
  \bibinfo{author}{\bibfnamefont{J.}~\bibnamefont{Martin}},
  \bibinfo{author}{\bibfnamefont{N.}~\bibnamefont{Rakotomalala}},
  \bibnamefont{and} \bibinfo{author}{\bibfnamefont{D.}~\bibnamefont{Salin}},
  \bibinfo{journal}{Physical Review Letters} \textbf{\bibinfo{volume}{79}},
  \bibinfo{pages}{5254} (\bibinfo{year}{1997}), ISSN \bibinfo{issn}{0031-9007,
  1079-7114},
  \urlprefix\url{https://link.aps.org/doi/10.1103/PhysRevLett.79.5254}.

\bibitem[{\citenamefont{Jinwoo and Tanaka}(2015)}]{jinwoo_local_2015}
\bibinfo{author}{\bibfnamefont{L.}~\bibnamefont{Jinwoo}} \bibnamefont{and}
  \bibinfo{author}{\bibfnamefont{H.}~\bibnamefont{Tanaka}},
  \bibinfo{journal}{Scientific Reports} \textbf{\bibinfo{volume}{5}}
  (\bibinfo{year}{2015}), ISSN \bibinfo{issn}{2045-2322},
  \urlprefix\url{http://www.nature.com/articles/srep07832}.

\bibitem[{\citenamefont{Truzzolillo and
  Cipelletti}(2018)}]{truzzolillo_hydrodynamic_2018}
\bibinfo{author}{\bibfnamefont{D.}~\bibnamefont{Truzzolillo}} \bibnamefont{and}
  \bibinfo{author}{\bibfnamefont{L.}~\bibnamefont{Cipelletti}},
  \bibinfo{journal}{Journal of Physics: Condensed Matter}
  \textbf{\bibinfo{volume}{30}}, \bibinfo{pages}{033001}
  (\bibinfo{year}{2018}), ISSN \bibinfo{issn}{0953-8984, 1361-648X},
  \urlprefix\url{http://stacks.iop.org/0953-8984/30/i=3/a=033001?key=crossref.0c1fb5f4b7abf9780e3d684ba5af76c7}.

\bibitem[{\citenamefont{Kheniene and Vorobev}(2013)}]{kheniene_linear_2013}
\bibinfo{author}{\bibfnamefont{A.}~\bibnamefont{Kheniene}} \bibnamefont{and}
  \bibinfo{author}{\bibfnamefont{A.}~\bibnamefont{Vorobev}},
  \bibinfo{journal}{Physical Review E} \textbf{\bibinfo{volume}{88}},
  \bibinfo{pages}{022404} (\bibinfo{year}{2013}), ISSN \bibinfo{issn}{1539-3755, 1550-2376},
  \urlprefix\url{https://link.aps.org/doi/10.1103/PhysRevE.88.022404}.

\bibitem[{\citenamefont{D'Errico
  et~al.}(2004{\natexlab{b}})\citenamefont{D'Errico, Ortona, Capuano, and
  Vitagliano}}]{derrico_diffusion_2004}
\bibinfo{author}{\bibfnamefont{G.}~\bibnamefont{D'Errico}},
  \bibinfo{author}{\bibfnamefont{O.}~\bibnamefont{Ortona}},
  \bibinfo{author}{\bibfnamefont{F.}~\bibnamefont{Capuano}}, \bibnamefont{and}
  \bibinfo{author}{\bibfnamefont{V.}~\bibnamefont{Vitagliano}},
  \bibinfo{journal}{Journal of Chemical \& Engineering Data}
  \textbf{\bibinfo{volume}{49}}, \bibinfo{pages}{1665}
  (\bibinfo{year}{2004}{\natexlab{b}}), ISSN \bibinfo{issn}{0021-9568,
  1520-5134}, \urlprefix\url{https://pubs.acs.org/doi/10.1021/je049917u}.

\bibitem[{\citenamefont{Barton}(1990)}]{barton_handbook_1990}
\bibinfo{author}{\bibfnamefont{A.~F.~M.} \bibnamefont{Barton}},
  \emph{\bibinfo{title}{Handbook of {Poylmer}-{Liquid} {Interaction}
  {Parameters} and {Solubility} {Parameters}}} (\bibinfo{publisher}{CRC Press},
  \bibinfo{year}{1990}).

\bibitem[{\citenamefont{Walderhaug and
  Knudsen}(2012)}]{walderhaug_aqueous_2012}
\bibinfo{author}{\bibfnamefont{H.}~\bibnamefont{Walderhaug}} \bibnamefont{and}
  \bibinfo{author}{\bibfnamefont{K.~D.} \bibnamefont{Knudsen}},
  \bibinfo{journal}{Journal of Solution Chemistry}
  \textbf{\bibinfo{volume}{41}}, \bibinfo{pages}{367} (\bibinfo{year}{2012}),
  ISSN \bibinfo{issn}{0095-9782, 1572-8927},
  \urlprefix\url{http://link.springer.com/10.1007/s10953-012-9796-5}.

\bibitem[{\citenamefont{Kuang et~al.}(2003)\citenamefont{Kuang, Maxworthy, and
  Petitjeans}}]{kuang_miscible_2003}
\bibinfo{author}{\bibfnamefont{J.}~\bibnamefont{Kuang}},
  \bibinfo{author}{\bibfnamefont{T.}~\bibnamefont{Maxworthy}},
  \bibnamefont{and}
  \bibinfo{author}{\bibfnamefont{P.}~\bibnamefont{Petitjeans}},
  \bibinfo{journal}{European Journal of Mechanics - B/Fluids}
  \textbf{\bibinfo{volume}{22}}, \bibinfo{pages}{271} (\bibinfo{year}{2003}),
  ISSN \bibinfo{issn}{09977546},
  \urlprefix\url{https://linkinghub.elsevier.com/retrieve/pii/S0997754603000359}.

\bibitem[{\citenamefont{Morra and Yuen}(2008)}]{morra_role_2008}
\bibinfo{author}{\bibfnamefont{G.}~\bibnamefont{Morra}} \bibnamefont{and}
  \bibinfo{author}{\bibfnamefont{D.~A.} \bibnamefont{Yuen}},
  \bibinfo{journal}{Geophysical Research Letters}
  \textbf{\bibinfo{volume}{35}}, \bibinfo{pages}{n/a} (\bibinfo{year}{2008}),
  ISSN \bibinfo{issn}{00948276},
  \urlprefix\url{http://doi.wiley.com/10.1029/2007GL032860}.

\bibitem[{\citenamefont{Mungall}(1994)}]{mungall_interfacial_1994}
\bibinfo{author}{\bibfnamefont{J.~E.}~\bibnamefont{Mungall}},
  \bibinfo{journal}{Physical Review Letters} \textbf{\bibinfo{volume}{73}},
  \bibinfo{pages}{288} (\bibinfo{year}{1994}).

\bibitem[{\citenamefont{Hansen}(2000)}]{hansen_hansen_2000}
\bibinfo{author}{\bibfnamefont{C.~M.} \bibnamefont{Hansen}},
  \emph{\bibinfo{title}{Hansen solubility parameters - {A} user's handbook}}
  (\bibinfo{publisher}{CRC Press LLC}, \bibinfo{address}{United States of
  America}, \bibinfo{year}{2000}).

\end{thebibliography}

\end{document}


\title{Supplemental material to\\Ultra-low effective interfacial tension between miscible molecular fluids}

\author{Alessandro~Carbonaro}
\email{alessandro.carbonaro@umontpellier.fr}
\affiliation{Laboratoire Charles Coulomb (L2C), UMR 5221 CNRS-Universit\'{e} de Montpellier,
Montpellier, France}
\author{Luca~Cipelletti}
\affiliation{Laboratoire Charles Coulomb (L2C), UMR 5221 CNRS-Universit\'{e} de Montpellier,
Montpellier, France}
\author{Domenico~Truzzolillo}
\affiliation{Laboratoire Charles Coulomb (L2C), UMR 5221 CNRS-Universit\'{e} de Montpellier,
Montpellier, France}

\null  
\nointerlineskip  
\vspace{5cm}
\let\snewpage \newpage
\let\newpage \relax
\maketitle
\let \newpage \snewpage
\break 
We provide here details on \textit{i)} the statistical significance of the water concentration dependence of the deformation velocity of drops in a miscible background, \textit{ii)} the simulations of the recirculating flow on top of spinning drops, and \textit{iii)} the limitations of the direct visualization of the radial Saffman-Taylor instability as a method to measure the effective interfacial tension (EIT) in water-glycerol systems.

\section{Statistical significance of the $c_w$ dependence of the deformation velocity}
Figure 4b of the main text shows the deformation velocity $v$ of drops evolving towards a dumbbell shape, as a function of the centripetal forcing $\Delta P_\omega$ and for different values of the water mass fraction $c_w$ of the drop. As seen in the figure, data may be divided in two sets: data points for $c_w\leq 0.75$ seem to be compatible with a straight line through the origin, while data for $c_w \geq 0.9$ apparently are not, since a linear fit exhibits a negative intercept with the $v$ axis. Since the intercept depends on $c_w$, hence possibly on the EIT, it is important to perform a statistical analysis of the data in Fig. 4b, in order to ascertain whether the difference between the two sets of data is statistically significant or is just due to experimental noise.

Due to measurement uncertainties, experimental data $y_{ex}$ measured as a function of a control parameter $x_{ex}$ and supposedly described by some physical law $y = f(x)$ will always be scattered around the model prediction. Assuming that data scattering results from a large number of mutually independent error sources, the underlying distribution of the residues $y_{ex} - f(x_{ex})$ may be assumed to be Gaussian \cite{m_r_spiegel_statistics_2008, press_numerical_1996}. 
When the number $N$ of experimental data points is small, typically $N\lesssim30$, statistics will deviate from that of large population samples. In particular, for small samples the Gaussian distribution of the residues will be replaced by the Student's $t$-distribution.

For our data, we consider the affine model $v= A\Delta P_{\omega} + B$ (Eq. (5) of the main text) and focus on the statistics of $\xi = v - A\Delta P_{\omega}$, where $A$ is obtained by simultaneously fitting the two sets of data (experiments with $c_w\leq 0.75$ and $c_w\geq 0.9$, respectively), imposing the same slope for both data sets. We denote respectively by $B_i$, $s_{i}$, $\mu_{i}$ and $N_i$ the mean, variance, expected value and number of samples for $\xi_i$, where the index $i = 1, 2$ refers to the two data sets. All values of the relevant experimental parameters are recapitulated in Table~SMT1.

\begin{table}
\begin{ruledtabular}
\begin{tabular}{ccccc}
 Data set&$c_w$&$N_i$&$B_i$ (m/s)&$s_i$ (m/s)\\ \hline
 $i=1$&$\leq 0.75$&9&$-5.4 \times 10^{-8}$&$3.3 \times 10^{-7}$ \\
 $i=2$&$\geq 0.9$&8&$-1.6 \times 10^{-6}$&$6.1 \times 10^{-7}$ \\
\end{tabular}\\\vspace{0.5cm}
TABLE SMT1. Parameters for the analysis of the statistical significance of the data of Fig. 4 of the main manuscript. See text for more details.
\end{ruledtabular}
\end{table}

To test the statistical significance of the difference between $B_1$ and $B_2$, we quantify the probability that the underlying expected values are actually identical, $\mu_{1}= \mu_{2}$, while $B_1$ and $B_2$ differ just by chance, having being obtained from a limited number of experimental data points. For small populations, the standardized difference between the two mean values, $t$, is distributed according to a Student's $t$-distribution with $N_1+N_2-2$ degrees of freedom, with
\begin{equation}
t = \frac{B_1 - B_2}{\sigma \sqrt{\sfrac{1}{N_1}+\sfrac{1}{N_2}}}\,,
\end{equation}
and where $\sigma$ is obtained from the experimental sample variances $s_{1}$ and $s_{2}$ as $\sigma = \sqrt{\frac{N_1 s_{1}^2 + N_2 s_{2}^2}{N_1 + N_2 -2}}$. Using the values reported in Table SMT1, we find $t = 6.43$, larger than $t = 4.073$, the edge of the $0.1\%$ confidence interval for a two-tailed $t$-distribution with $N_1+N_2-2 =15$ degrees of freedom. We conclude that $\mu_{1} \neq \mu_{2}$ with probability larger than 99.9\%, i.e. that the difference between $B_1$ and $B_2$ is indeed statistically significant.

\section{Recirculating secondary flow and dependence on $\Delta P_\omega$}
In any spinning drop tensiometry experiment, the background fluid is not at rest, but rather flows towards the axis of the capillary inducing an extra normal stress on the drop surface \cite{currie_buoyancy_1982}. This phenomenon, present even for stationary, ellipsoidal drops, has been invoked to rationalize the origin of dumbbell shapes when the interfacial tension between the drop and the background fluid is extremely weak~\cite{manning_interfacial_1977, zoltowski_evidence_2007}. Indeed, in a spinning capillary the centripetal pressure $P$ depends on the axial, $x$, and radial, $r$, coordinates as:
\begin{equation}
\label{eq:P_centripetal}
P(r,x) = \int_0^r  \rho(r',x) \omega^2 r' \mathrm{d}r'  \,,
\end{equation}
where $\omega$ is the rotational speed and $\rho$ the fluid-dependent mass density. Since the mass density of the drop is lower than that of the background fluid, for a given distance $r$ from the capillary axis $P$ is lower in the $x$ region occupied by the drop, as shown in Fig. SM1, which displays the axial profile of $P$ for a cylindrical drop capped by two hemispheres. The horizontal pressure gradient due to the pressure drop in the region  $x_A<x<x_D$ induces a recirculation flow in the capillary.
\begin{figure}[h]
	\centering
	\includegraphics[width = 0.8\textwidth]{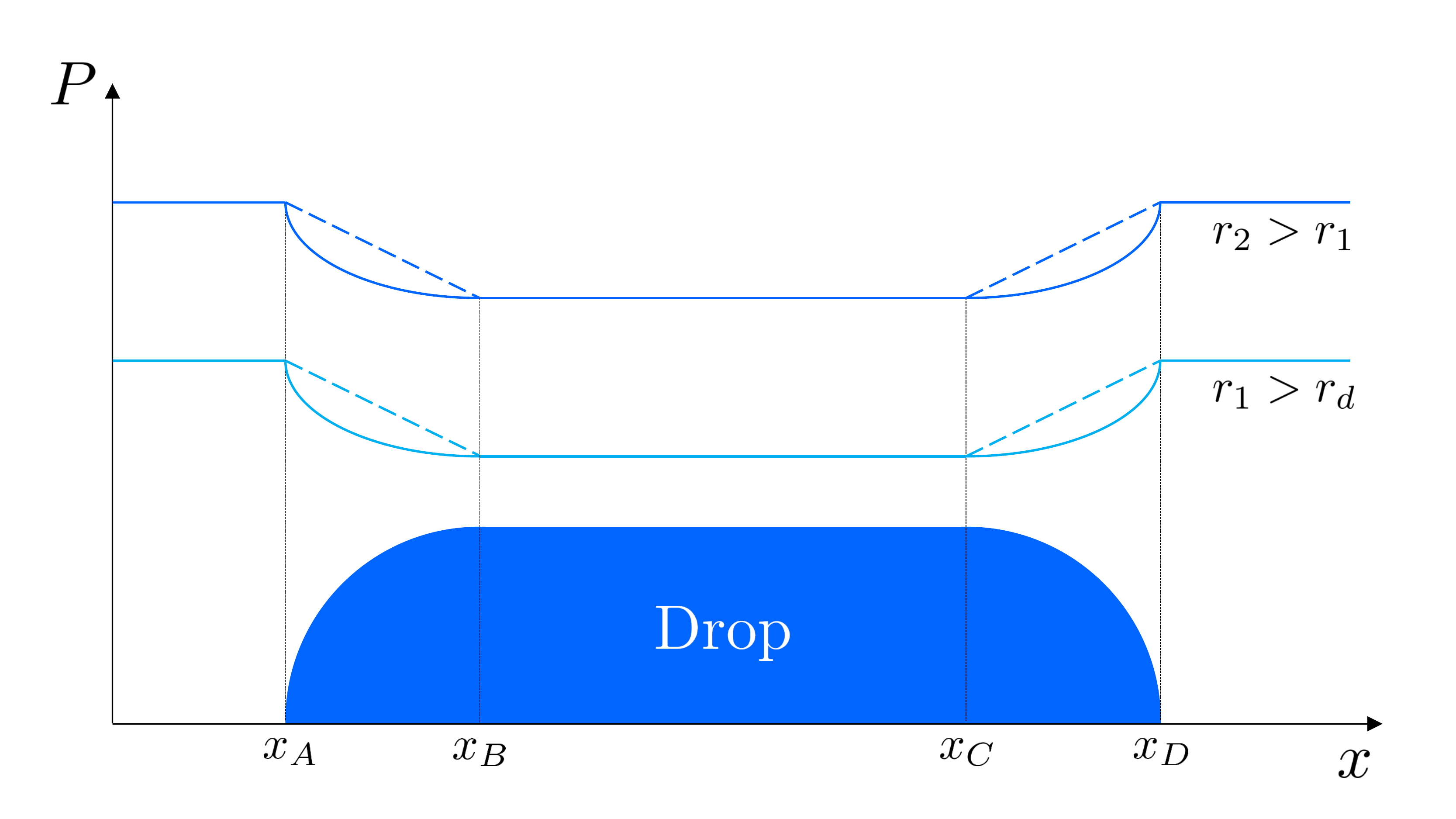}\\
	\textbf{Figure SM1: }Axial dependence of the pressure in the background fluid in a spinning drop experiment, for two distances $r$ from the capillary axis, both larger than the drop radius. The drop shape is shown as the blue region. The solid lines show the pressure profile for a cylindrical drop capped by two hemispheres. The dashed lines show the simplified, piece-wise linear profile used in the calculations. The decrease of $P$ in the $x$ range occupied by the drop is responsible for the recirculation flow of the background fluid shown schematically in Fig. 5 of the main manuscript.
	\label{schemeP_supmat}
\end{figure}

To gain insight on this flow, we solve numerically the Navier-Stokes equations for the background fluid in the region $x_A  \leq x \leq x_D$ and $r_d \leq r \leq r_C$, with $r_d$ and $r_C$ the radius of the drop and of the capillary, respectively. We assume all fluids to be incompressible and write the momentum conservation equation by splitting the pressure into its steady and unsteady contributions: $p(x,r) = P(x,r) + p'(x,r)$, where the r.h.s. terms represent the inhomogeneous centripetal pressure and the pressure associated to fluid motion, respectively. One obtains
\begin{equation}
\frac{\partial \vec{v}}{\partial t} + (\vec{v} \cdot \nabla) \vec{v} = -\frac{1}{\rho}\nabla (P+p') + \nu \nabla^2 \vec{v} \,,
\end{equation}
where $\vec{v}$, $\rho$ and $\nu$ are the fluid velocity, density and kinematic viscosity, respectively.
Separating the two contributions to the pressure field is equivalent to including in the momentum equation a source term arising from the (steady) pressure gradient, $-\frac{1}{\rho}\nabla P = \vec{F}$:
\begin{equation}
	\frac{\partial \vec{v}}{\partial t} + (\vec{v} \cdot \nabla) \vec{v} = -\frac{1}{\rho}\nabla p' + \nu \nabla^2 \vec{v} + \vec{F}
\end{equation}

The cylindrical symmetry of the capillary allows the problem to be solved in two dimensions, yielding the following set of equations:
\begin{equation}\label{momentum_x}
	\frac{\partial u}{\partial t} + u \frac{\partial u}{\partial x} + v \frac{\partial u}{\partial r} = -\frac{1}{\rho}\frac{\partial p'}{\partial x} + \nu \left(\frac{\partial^2 u}{\partial x^2} + \frac{\partial^2 u}{\partial r^2}\right)+ F_x
\end{equation}
\begin{equation}\label{momentum_y}
\frac{\partial v}{\partial t} + u \frac{\partial v}{\partial x} + v \frac{\partial v}{\partial r} = -\frac{1}{\rho}\frac{\partial p'}{\partial r} + \nu \left(\frac{\partial^2 v}{\partial x^2} + \frac{\partial^2 v}{\partial r^2}\right)+ F_r
\end{equation}
\begin{equation}\label{Poisson_pressure}
	\frac{\partial^2 p'}{\partial x^2} + \frac{\partial^2 p'}{\partial r^2} = -\rho \left[\left(\frac{\partial u}{\partial x}\right)^2 + 2 \frac{\partial u}{\partial r}\frac{\partial v}{\partial x} + \left(\frac{\partial v}{\partial r}\right)^2\right]
\end{equation}
where $u$ and $v$ denote the $x$ and $r$ components of $\vec{v}$, respectively.
Equation~\ref{Poisson_pressure} is the Poisson equation for the pressure obtained by inserting the incompressibility condition for the background fluid ($\nabla \cdot \vec{v} = 0$) in the momentum equation.

We numerically solve Eqs.(\ref{momentum_x}-\ref{Poisson_pressure}) for the background fluid ($r> r_{d}$), in the $x$ range occupied by the drop, with boundary conditions $(u, v)= (0,0)$ and $\frac{\partial p}{\partial \vec{n}} = 0$ at the drop-background fluid interface and at the capillary wall, where $\vec{n}$ is the unit vector normal to each boundary. The solution for $u$, $v$ and $p$ is obtained by performing a central discretization on the variables~\cite{quarteroni_numerical_2017} and by iterating until stability is reached. Note that we are interested in understanding the general behavior of the recirculating flow rather than its detailed behavior. Accordingly, for the sake of simplicity we approximate the pressure profile $P(x)$ by a piece-wise linear function with a trapezoidal shape and a pressure drop $\Delta P_\omega = \frac{\omega^2}{2}( \rho_f - \rho_d) r_d^2 $ in correspondence with the drop body (see dashed lines in Fig. SM1).

\begin{figure}
	\centering
	\subfloat[\label{Poiseille}]{\includegraphics[width = 0.5\linewidth]{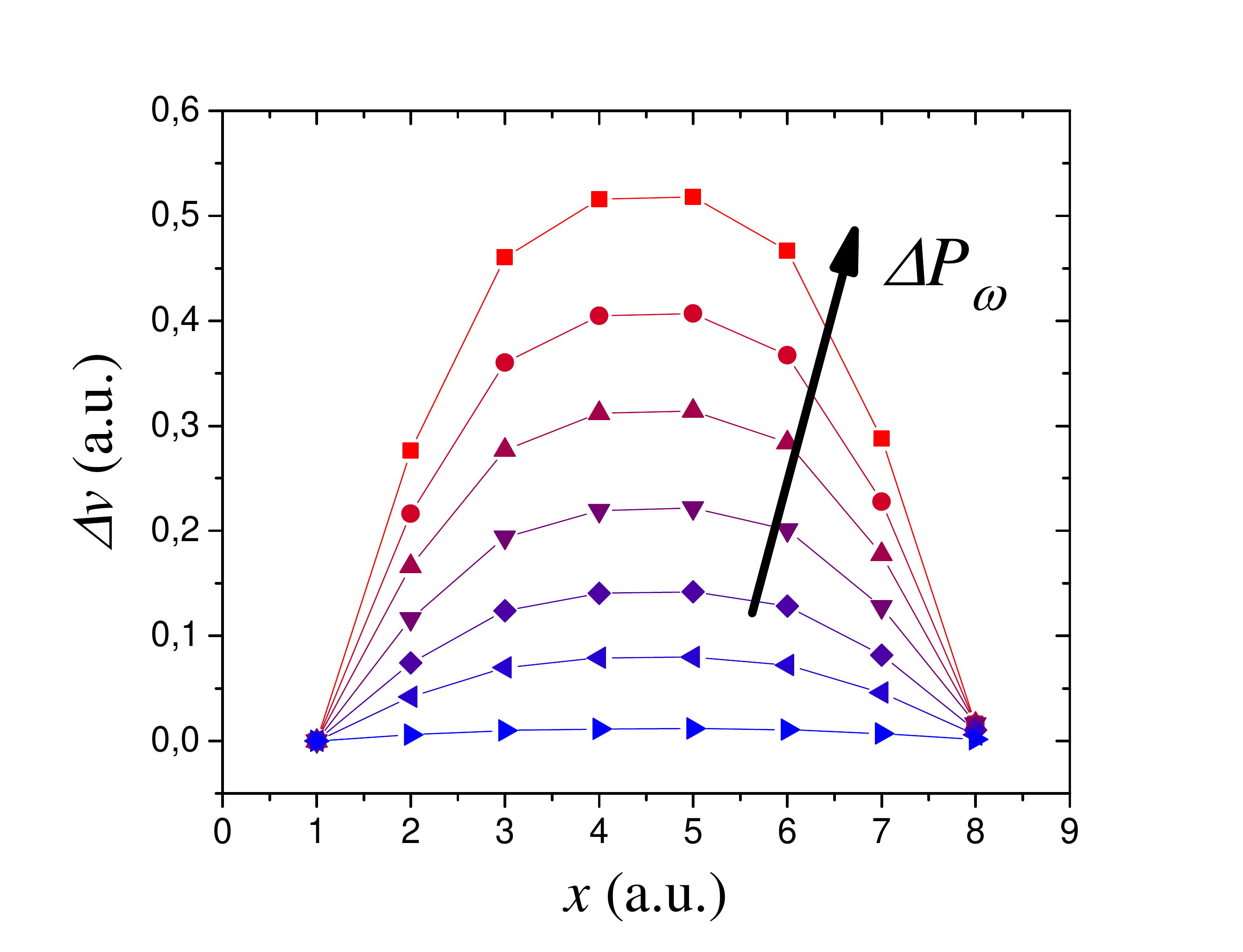}}
	\hfill
	\subfloat[\label{v_forcing}]{\includegraphics[width = 0.5\linewidth]{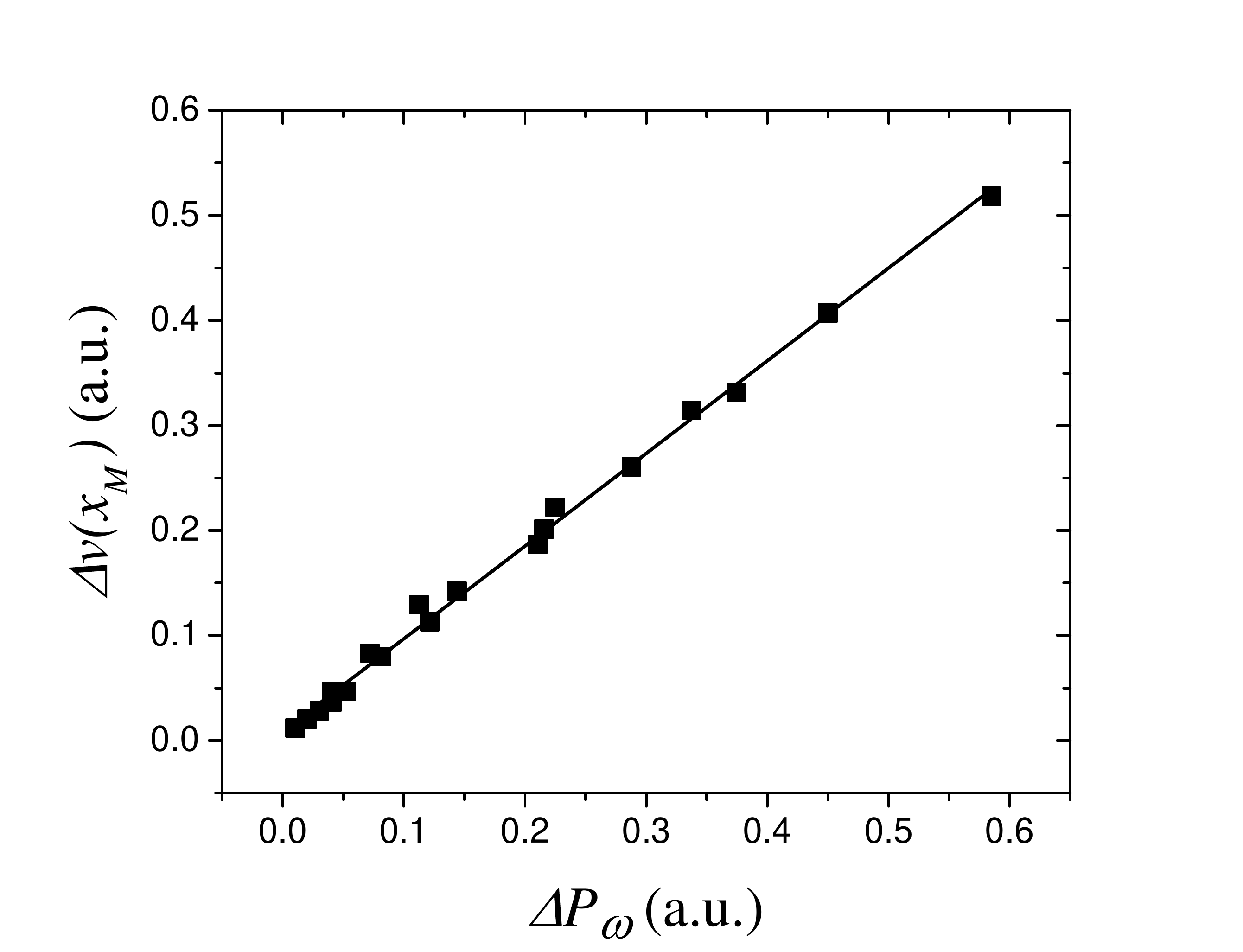}}\\
	\textbf{Figure SM2:} (a) Radial velocity difference $\Delta v(x) = v(x_B)-v(x)$ of the background fluid for several magnitudes of the centripetal forcing $\Delta P_\omega$, obtained from a numerical solution of Eqs.(\ref{momentum_x}-\ref{Poisson_pressure}), with $x_B = 1$, $x_C = 8$, $r_d = 15$ and $r = 19$. $\Delta P_\omega$ varies from $0.01$ to $0.58$ by steps of $0.095$, from the bottom curve to the top one. (b) Symbols: maximum value of $\Delta v$, attained at $x_M = 4.5$, as a function of $\Delta P_\omega$. The line is a linear fit through the origin.	
\end{figure}

In order to investigate the external forcing on the drop surface due to the recirculating flow, we evaluate the radial component of the velocity in the $x$ range occupied by the drop body, $x_B<x<x_C$. Since we are interested in the radial deformation of the drop towards a dumbbell shape, we look for a difference between $v$ at the center of the drop and close to the heads. Figure SM2 a) shows the axial dependence of $\Delta v (x) = v(x_B)-v(x)$, for various $\Delta P_\omega$. The data are obtained for $r= 19$, but they are representative of all $r$ within the gap between the drop and the capillary wall. Clearly, the magnitude of the radial velocity is larger at the center of the drop than near the heads, resulting in the drop surface being pushed towards the capillary axis stronger at its center than at the heads, consistent with the development of a dumbbell shape. Figure SM2 a) shows also that the velocity imbalance increases with increasing $\Delta P_\omega$, i.e. with increasing angular velocity or density mismatch. To quantify the dependence on the centripetal forcing, we plot in Fig. SM2 b) the maximum value of $\Delta v$, attained in correspondence to the center $x_M$ of the drop body, as a function of $\Delta P_\omega$. The magnitude of the recirculating flow, $\Delta v(x_M)$, is found to depend linearly on $\Delta P_\omega$. This justifies writing the normal stress induced by the recirculating flow at the onset of the instability as $n_E = \alpha \Delta P_\omega$, with $\alpha$ a positive constant, as it was done in Sec. III A of the main text. Moreover, this explains why the experimental deformation velocity depends linearly on $\Delta P_\omega$, as seen in Fig. 4b of the main text.

\section{Can the direct visualization of the radial Saffman-Taylor instability be used as a method to measure $\Gamma_e$ in water-glycerol systems?}
The Saffman-Taylor (S-T) instability occurs when a less viscous fluid displaces a more viscous one confined in a porus medium. The boundary between the fluids is uneven and the less viscous fluid develops fingers protruding into the more viscous one. Most experimental research on viscous fingering has been performed using Hele-Shaw cells, which consist of two closely spaced, parallel plates of glass. In the widely used radial configuration, the more viscous fluid is confined between the two plates and the less viscous fluid is injected through a small hole drilled in the center of one of the plates. We have recently shown that the analysis of the characteristic wavelength  $\lambda_{ST}$ at the onset of the radial S-T instability is a valuable tool to investigate interfacial stresses in miscible colloidal and polymeric fluids~\cite{truzzolillo_off-equilibrium_2014,truzzolillo_nonequilibrium_2016}. One may thus wonder if the same strategy could be employed to measure $\Gamma_e$ for water-glycerol systems.

Various experiments, however, have suggested that radial S-T instabilities occurring when water displaces glycerol \cite{paterson_fingering_1985} ---or more generally when two different water-glycerol mixtures are used to generate S-T patterns \cite{bischofberger_fingering_2014}--- are not affected by capillary forces. In 1985, Paterson \cite{paterson_fingering_1985} showed that in the absence of interfacial tension a cut-off wavelength characterizes the onset of the S-T instability. Namely, his theoretical analysis shows that for $\Gamma_e = 0$ the growth of the instability is dominated by viscous dissipation, giving rise to a characteristic wavelength dictated only by the cell gap $b$: $\lambda_{ST}\simeq 4b$. By contrast, for $\Gamma_e>0$, the wavelength characterizing the onset of the instability depends not only on $b$, but also on the injection rate of the less viscous fluid, the viscosity contrast between the fluids and the interfacial tension~\cite{truzzolillo_off-equilibrium_2014}. In particular, all other parameters being fixed, $\lambda_{ST}$ grows with $\Gamma_e$, such that $\lambda_{ST}\simeq 4b$ is the minimum wavelength of a S-T instability observable in a radial viscous fingering experiment. The experimental data for water/glycerol systems of Refs.~\cite{paterson_fingering_1985,bischofberger_fingering_2014} are consistent with $\Gamma_e \simeq 0$, since it was found that $\lambda_{ST}\simeq 4b$.

In order to evaluate the feasibility of measuring $\Gamma_e$ in water-glycerol systems via the visualization of S-T patterns, we perform numerical simulations based on the linear analysis introduced by Miranda and Widom~\cite{miranda_radial_1998}. Our goal is to quantify the minimum value of $\Gamma_e$ that could be measured by this method in a Hele-Shaw experiment with optimized yet realistic parameters. In the simulations, the fluid-fluid interface is supposed to expand radially. A perturbation around this circular interface develops with time, due to the instability. It is convenient to decompose the perturbation in Fourier modes with complex amplitude $\zeta_n(t)$, $n = 1, 2,...$. Assuming that the noise giving rise to the instability is a complex number $\zeta_n^0$, with a random phase and a $n$-independent modulus, the time evolution of the amplitude of the $n$-th mode is given by~\cite{miranda_radial_1998,truzzolillo_off-equilibrium_2014}:
\begin{equation}\label{zeta}
\zeta_n(t)=\zeta_n^0\left\{\left(K(t)\frac{(nA-1)}{n^2(n-1)}^{nA-1}\right)\exp\left[(nA-1)\left(\frac{1}{K(t)}\frac{n(n^2-1)}{nA-1}-1\right)\right]\right\} \,.
\end{equation}
In Eq.~(\ref{zeta}), $A=(\eta_2-\eta_1)/(\eta_2+\eta_1)>0$ is the viscosity contrast between the two fluids and $K(t)=[r(t)Q]/(2\pi\beta)$, where $r(t)$ is the distance of the unperturbed fluid-fluid interface from the center of the cell, $Q$ is the area covered by the injected fluid per unit time, and $\beta=b^2\Gamma/[12(\eta_1+\eta_2)]$, with $b$ the cell gap and $\Gamma$ the interfacial tension between the two fluids. Note that Eq.~(\ref{zeta}) only holds for $nA >1$, a condition fulfilled for water displacing glycerol for all $n > 1$. We emphasize that Eq.~(\ref{zeta}) is derived in the framework of a quasi 2D theory that neglects the curvature of the interface along the direction perpendicular to the glass plates. As a result, this theory poses no lower bound on the wavelength of the instability: values of $\lambda_{ST} < 4b$ issued from the simulations should therefore be regarded as nonphysical and discarded.

\begin{figure}[htbp]
  \includegraphics[width=12cm]{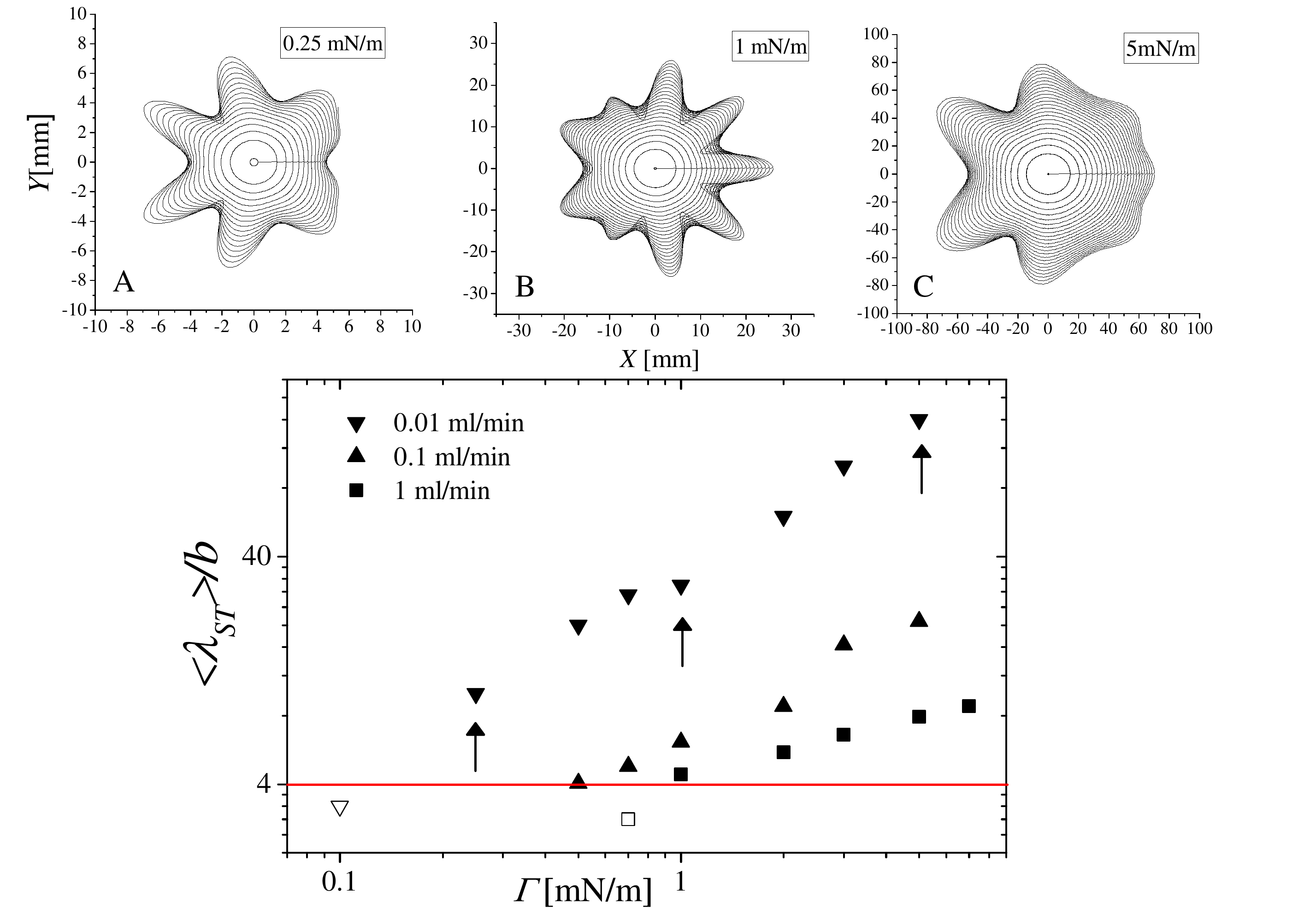}\\
  \textbf{Figure SM3.} Top row:  representative patterns obtained at $\dot{V}=0.01$ ml/min and for three values of the interfacial tension, as indicated by the labels. The curves display the interface between the two fluids at times that increase respectively from 10 s
by steps of 10 s (A), from $10^2$ s by steps of $10^2$ s (B) and from $10^3$ s by steps of $10^3$ s (C). Bottom panel: Average wavelength of the radial S-T instability $\langle\lambda_{ST}\rangle$ at its onset as a function of the interfacial tension $\Gamma$ for three different injection rates. $\langle\lambda_{ST}\rangle$ is obtained by averaging results from 10 simulation runs, with $b=250~\mu$m and a random noise amplitude $|\zeta_n^0|=10^{-9}$ m. The arrows indicate the data points corresponding to the three panels of the top row. Open symbols refer to nonphysical values $\langle\lambda_{ST}\rangle < 4b$ and should be discarded, as discussed in the text. \label{SF-panel}
\end{figure}

Using Eq.~(\ref{zeta}) and summing the contributions of 500 modes, we simulate the S-T instability in a Hele-Shaw cell with $b=250$ $\mu \mathrm{m}$, as in previous works on miscible fluids \cite{bischofberger_fingering_2014,truzzolillo_nonequilibrium_2016}. The top row of Fig. SM3 shows three representative sets of interface positions at various times, obtained from simulations with different values of $\Gamma_e$, as indicated by the labels. We measure the wavelength of the instability at its onset by counting the number of nodes (inflection points) along the first contour line for which inflection points clearly emerge from the background noise. The wavelength thus obtained is plotted in the bottom panel of Fig. SM3, for three injection rates $\dot{V}$ within the typical range of experimentally accessible values~\cite{bischofberger_fingering_2014,truzzolillo_off-equilibrium_2014,truzzolillo_nonequilibrium_2016}. Keeping in mind that data points with $\langle\lambda_{ST}\rangle < 4b$ are nonphysical, one concludes that for the parameters used here $\Gamma_e$ values smaller than about 0.1 mN/m are unaccessible. This sets an upper bound for the interfacial tension between water and glycerol measured in previous Hele-Shaw experiments. Note that, in principle, thinner Hele-Shaw cells would allow smaller values of $\lambda_{ST}$ and hence of $\Gamma_e$ to be measured. However, for $b<250$ $\mu \mathrm{m}$ it is hard to pre-fill homogeneously the cell with the more viscous fluid (glycerol in our case), because of the heterogeneous local wetting of the fluid on the cell walls, which are unavoidably irregular on the micron scale. This typically gives rise to an anisotropic distribution of the fluid in the cell, leading to erratic experimental results.

\begin{figure}[htbp]
  \includegraphics[width=12cm]{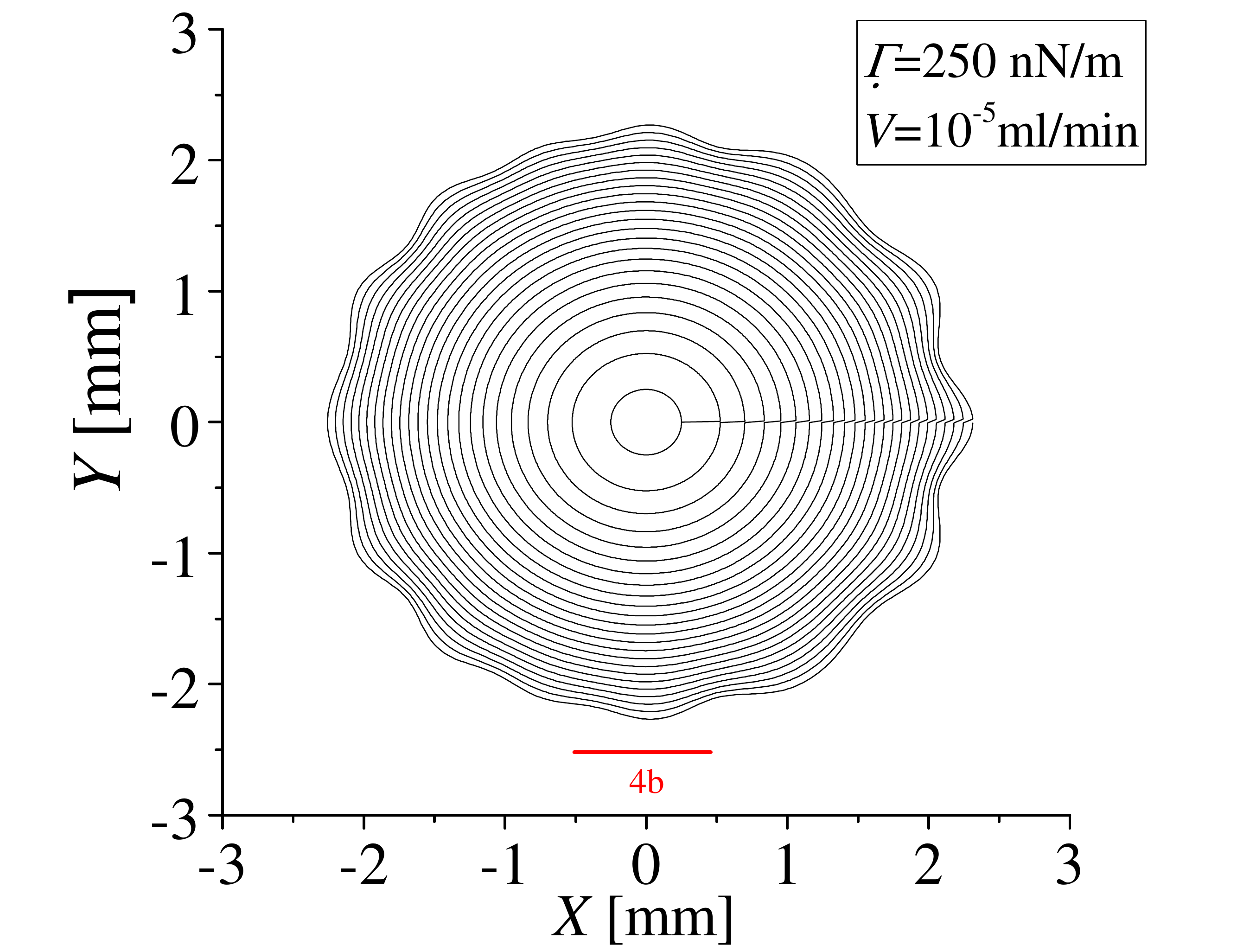}\\
  \textbf{Figure SM4.} Typical S-T pattern obtained by imposing $\Gamma_e=250 \times 10^{-6}~\mathrm{mN}/\mathrm{m}$, as in our spinning drop experiments. The injection rate ($10^{-5}$ ml/min) has been chosen to match the limiting condition $\langle\lambda_{ST}\rangle \approx 4b$. The time needed to the interface to reach the radius at which the instability becomes measurable is $t_{onset}\simeq 314$ s. \label{SF-250nN}
\end{figure}

Another strategy to access very small $\Gamma_e$ values by investigating the S-T instability might consist in using extremely low injection rates. Indeed, the bottom graph of Fig. SM3 shows that reducing the injection rate shifts the lower bound on the measurable EIT to lower values. To test whether this approach would be practically feasible, we run simulations imposing $\Gamma_e=250 \times 10^{-6}~\mathrm{mN}/\mathrm{m}$, the value inferred from our spinning drop experiments. We tune the injection rate so as to keep the wavelength of the perturbation at the onset of the instability just above the limiting value $\lambda_{ST} = 4b$. The resulting pattern is shown in Fig. SM4. We find that an extremely small injection rate would be required: $\dot{V}=10^{-5}$ ml/min. Even more importantly, the onset of the instability would occur 314 s after starting the injection, a time span one order of magnitude larger than that in our spinning drop experiments. Over such a long time, diffusion would significantly smear out the interface, reducing the EIT well below the resolution of any method currently available. Therefore, the measurement of $\Gamma_e$ through the detection of S-T patterns in water-glycerol systems appears to be unfeasible, because diffusion smears the interface at very low injection rates, while viscous dissipation alone dictates the interface shape at higher rates.
